\documentclass[english,aps,nofootinbib,pra,notitlepage,reprint,floatfix]{revtex4-1}
\pdfoutput=1
\usepackage[T1]{fontenc}
\usepackage[latin9]{inputenc}
\usepackage{color}
\usepackage{babel}

\usepackage{amsmath}
\usepackage{graphicx}
\usepackage{subfig}
\usepackage{amssymb}
\usepackage[pdftex,unicode=true, pdfusetitle,
 bookmarks=true,
 breaklinks=false,pdfborder={0 0 0},backref=true,colorlinks=false, hyperindex=true]
 {hyperref}

\makeatletter

\providecommand{\tabularnewline}{\\}

\@ifundefined{textcolor}{}
{%
 \definecolor{BLACK}{gray}{0}
 \definecolor{WHITE}{gray}{1}
 \definecolor{RED}{rgb}{1,0,0}
 \definecolor{GREEN}{rgb}{0,1,0}
 \definecolor{BLUE}{rgb}{0,0,1}
 \definecolor{CYAN}{cmyk}{1,0,0,0}
 \definecolor{MAGENTA}{cmyk}{0,1,0,0}
 \definecolor{YELLOW}{cmyk}{0,0,1,0}
 }

\makeatother

\begin{document}

\preprint{Working draft: Not for distribution}

\title{Effect of Interatomic Separation on Entanglement Dynamics in a Two-Atom Two-Mode Model}

\author{K. Sinha}

\email{kanu@umd.edu}

\author{N. I. Cummings}

\email{nickc@umd.edu}

\author{B. L. Hu}

\email{blhu@umd.edu}

\affiliation{Joint Quantum Institute and Department of Physics\\
University of Maryland, College Park, Maryland 20742-4111, USA}

\date{August 6, 2011}

\begin{abstract}
We analyze the time evolution of quantum entanglement in a model consisting of two two-level atoms interacting with a two-mode electromagnetic field for a variety of initial states and interatomic separations.  We study two specific atomic separations which give rise to symmetric atom-field couplings.  For general atomic distances we consider a subset of initial states analytically, and then treat the more general situation numerically. We examine a variety of qualitative features such as entanglement sudden death, dynamical generation, protection, and transfer between subsystems.  Our analysis shows a stark contrast in features of entanglement between the two special coupling schemes often considered; further, these features are uncharacteristic of those arising for general distances, due to the high degree of symmetry present in the special cases.  The variety of behaviors in these two-mode cases suggest the importance of considering atomic separation carefully for any model where two atoms interact with a common field.
\end{abstract}

\maketitle
\global\long\def\ket#1{\left|#1\right\rangle }
\global\long\def\bra#1{\left\langle #1\right|}
\global\long\def\tr{\mathrm{Tr}}
\global\long\def\op#1{\hat{#1}}
\global\long\def\defeq{\equiv}
\global\long\def\vect#1{\boldsymbol{#1}}
\global\long\def\leftexp#1#2{{\vphantom{#2}}^{#1}{#2}}


\section{Introduction}

Quantum entanglement has been extensively studied, both due to its
fundamental significance in quantum theory \cite{BellIneq,EPR} and
its utility as a resource for quantum communication and quantum
information processing
\cite{NielsenChuangBook,BennettDiVincenzo2000}. Atomic physics offers
a domain with sufficient control of the system and isolation from
noise that it has been possible to perform precision experiments on
quantum entanglement \cite{OBrien2010,RaimondEtAl2001}, and,
therefore, it is also a productive target for theoretical study of
the issue as well.

One of the simplest scenarios for theoretical studies of the dynamics of entanglement
between atoms is that of two atoms which are isolated from one another
and interact with different electromagnetic fields. Studying this
sort of model led to the discovery of the entanglement sudden death
(SD) phenomenon \cite{YuEberly2004,YuEberly2009,Almeida2007,YonacYuEberly2006,YonacYuEberly2007},
in which entanglement decays to zero in a finite time rather than
asymptotically. SD has garnered significant interest because it is
both unintuitive and potentially undesirable.

An alternative simple model for atom-field interaction in which to
study entanglement dynamics is the model studied by Dicke \cite{Dicke} and Tavis and Cummings \cite{Cummings68},where one
assumes all atoms are grouped in a sample whose size is small
compared to the resonant wavelength (resulting in identical coupling
to every atom). However, such a simplistic model may miss the variety
of behavior that can result when the atoms are not confined to such a
small sample. Entanglement dynamics have been shown to have a
significant distance dependence when two atoms are interacting with a
common field \cite{TaneichiKobayashi2003}. For atoms weakly
interacting with a continuum of field modes in the Born-Markov
approximation, it has been shown
\cite{FicekTanas2008,TanasFicek2004a,TanasFicek2004b,FicekTanis2006}
that changing the atomic separation of two atoms can affect whether
there is SD and whether there is revival of entanglement, as well as
modify the dynamical generation of entanglement; in short, the
qualitative features are sensitive to the atomic spacing. At short
interatomic distance non-Markovian effects associated with induced
interactions between the atoms due to the quantum field become more
pronounced whose qualitative behavior varies greatly with different
classes of initial states, as detailed in the study of
\cite{ASH2006}. So it is clear that distance dependence gives rise to
a significant variety of behaviors, and some seemingly innocuous approximations 
can qualitatively alter the entanglement dynamics in unintuitive ways
\cite{ASH2009,JingLuFicek2009}.

In view of this, we seek to study entanglement dynamics in a model
that is sufficiently complex to manifest some of this variety of
behavior yet simple enough that the dynamics may be understood in
considerable detail and obtained with fewer approximations. The
simplest model one may pursue along this line would be two two-level
atoms coupled to a single mode with couplings that reflect the
distance dependence; however, as discussed in Sec.~\ref{sec:sysham}, one needs to include at least two field modes before
any non-trivial distance dependence can arise in the problem. So we will 
study such a model with two field modes here. We choose to have traveling wave modes so as to eliminate the effect of atomic center of mass position and focus on the relative separation between the atoms, this can be experimentally realized in the situation where one has two counter-propagating modes in a toroidal resonator cavity \cite{KimbleTorus,DanAlton}. The motivation for working with as few modes as as possible rather than a continuum is to explore the variety of dynamical behavior that can arise in a non-dissipative system and its sensitivity to initial state.

In order to better qualify the entanglement dynamics, we must specify a way of quantifying 
entanglement. For two qubits one may compute the entanglement 
of formation efficiently in terms of the Wootters concurrence, and the concurrence itself may be used as an entanglement monotone \cite{Wooters1998}.   This is the quantity we will use most often to quantify the atom-atom entanglement in this paper.
Alternatively, to quantify entanglement one can also use the negativity \cite{VidalWerner2002} (which quantifies the degree to which the state violates the Peres-Horodecki positive partial transpose condition).  This is useful on larger Hilbert spaces, and we use this to quantify entanglement when considering aspects of the the entanglement dynamics involving field states.

In section \ref{sec:sysham} we describe a useful way to parameterize the class of Hamiltonians that 
can give rise to different entanglement dynamics due to atomic separation.  
In section \ref{sec:Sp} we focus on two special cases in terms of separation with the aim of
illustrating the variety of different behaviors that can arise in different initial atomic and field states.
Specifically, we will show how SD, dynamical entanglement generation,
and other phenomena differ between these two cases by
comparing with analogous situations of the two
well-studied types of models - where each system
is a single two-level atom interacting with a single field
mode \cite{YonacYuEberly2006,YonacYuEberly2007,Chan2009} and two two-level atoms interacting with a single
common field mode \cite{KimKnight2003,TessierDeutsch2003}. We will
show that the entanglement dynamics in these models differs
significantly to illustrate the effect of atomic separation.  Finally, in section \ref{sec:Gen} we examine the entanglement dynamics 
for other atomic separations, using analytic results for a few cases and 
numerical results more generally, to examine to what degree the 
behaviors of the special cases examined generalize.  To look at the entanglement evolution for intermediate values of separation analytically, we use the resolvent expansion method of \cite{ASH2006} restricted to initial states with a single excitation among the atoms and field modes. Also, for the most general case we can look at the dynamics by numerical diagonalization of the Hamiltonian for all the other initial states and arbitrary separations. In this way we explore 
the implications of interatomic distance on entanglement dynamics in this 
simplest non-trivial context.

\section{System Hamiltonian\label{sec:sysham}}

Consider a pair of identical two-level atoms coupled via the minimal coupling 
interaction Hamiltonian 
to a collection of electromagnetic (EM) field modes
under the dipole and rotating-wave approximation.
We will assume that atoms can be effectively considered to have fixed center of mass positions $\vect R_{j}$.
The interaction Hamiltonian of the system in the Schr\"{o}dinger
picture has the form \begin{equation}
\op H_{I}=\hbar\sum_{j}\sum_{q}g_{j,q,s}\op{\sigma}_{j}^{+}\op a_{q,s}+g_{j,q,s}^{*}\op{\sigma}_{j}^{-}\op a_{q,s}^{\dagger},
\end{equation}
 where $\op{\sigma}_{j}^{+}$ and $\op{\sigma}_{j}^{-}$ are the raising and lowering operators for 
the two-level system representing the $j^{\textrm{th}}$ atom, and $\op a_{q,s}^{\dagger}$ and $\op 
a_{q,s}$ are the creation and annihilation operators corresponding to the $q^{\textrm{th}}$ normal 
mode with frequency $\omega_{q}$ and polarization indexed by $s$ in some mode decomposition of the EM field with 
(complex) classical electric field mode function $\vect f_{q,s}\left(\vect r\right)$.  
Let us assume that the atomic transition in question does not alter the angular momentum of the atom.  Under these assumptions 
\begin{equation}
g_{j,q}\defeq - i d_{j}\vect e_{d,j}\cdot\vect f_{q,s}\left(\vect R_{j}\right)\sqrt{\frac{\omega_0}{2 \hbar\omega_{q} \epsilon_{0}V}},
\end{equation}
where the field is quantized in a region of volume V, $d_{j}$ is the complex dipole matrix element for the transition in the $j^{\textrm{th}}$ atom, and the dipole moment has a direction $\vect{e}_{d,j}$ \cite{Anastopoulos00,DutraBook2004}.

In general, the mode functions $\vect{f}_{q,s}\left(\vect R_{j}\right)$ can be quite complicated, 
and even for a single atom there can be position dependence in the dynamics arising from the 
boundary conditions that the mode functions obey. In order to distinguish these effects from the 
effect of atomic separation, we will consider mode functions of the form 
\begin{equation}
f_{q}\left( u,v \right) e^{\pm i k_q w} \vect{\varepsilon}_{q,s} 
\end{equation}
in terms of some set of coordinates $\left( u,v,w \right)$, with $s=1,2$ indexing the polarization.  This form describes 
traveling-wave mode functions that are solutions to an EM boundary value problem which is translation invariant under translations in the coordinate $w$, and it appears in a number of interesting physical systems, including spherical resonators \cite{MabuchiKimble1994} and toroidal resonators \cite{AokiEtAl2006,DanAlton}.
We will consider situations in which all atoms share the same coordinates $u=u_0$ and $v=v_0$, differing only in the coordinate $w_j$, so that the variation in behavior due to the atomic separation will manifest without additional position dependence arising from the boundary conditions.  Let us further assume that the atoms are arranged such that the atomic transition dipole vectors are aligned with one of the mode polarizations, so $\vect{e}_{d,j} \cdot \vect{\varepsilon}_{q,s} = \delta_{s,1}$.  In this situation only one polarization is relevant, and all polarization labels will be suppressed.  With these further assumptions we may express the coupling constants as 
\begin{equation}
g_{j,q} = -i d_{j} f_{q}\left( u_0,v_0 \right) e^{\pm i k_q w} \sqrt{\frac{\omega_0}{2 \hbar\omega_{q} \epsilon_{0}V}};\label{eq:CouplingForq}
\end{equation}
however, it turns out that without loss of generality one may study a much smaller set of Hamiltonians. 
 
Given a particular set of complex phases for the coupling constants
$g_{j,q}$, one may make a trivial basis transformation \begin{equation}
\op U_{b}=\prod_{l}e^{-i\xi_{l}\op{\sigma}_{l}^{z}/2}\prod_{s}e^{i\zeta_{s}\op a_{s}^{\dagger}\op a_{s}},\end{equation}
 which simply amounts to redefining the reference for the phases of
the atoms by $\op{\sigma}_{j}^{+}\rightarrow\op{\sigma}_{j}^{+}e^{-i\xi_{j}}$
and for the fields by $\op a_{q}\rightarrow\op a_{q}e^{-i\zeta_{q}}$.
After this transformation, the interaction Hamiltonian becomes 
\begin{equation}
	\op H_{I}^{\prime}=\op U_{b}\op H_{I}\op U_{b}^{\dagger}=\hbar\sum_{j}\sum_{q}g_{j,q}^{\prime}\op{\sigma}_{j}^{+}\op a_{q}+\left(g_{j,q}^{\prime}\right)^{*}\op{\sigma}_{j}^{-}\op a_{q}^{\dagger}
\end{equation}
 with $g_{j,q}^{\prime}=g_{j,q}e^{-i\xi_{j}}e^{-i\zeta_{q}}$, 
and we may obtain the solution to the original problem by using
the transformed Hamiltonian: 
\begin{align}
	\ket{\psi\left(t\right)}= & e^{-i\op H_{I}t/\hbar}\ket{\psi\left(0\right)}=\op U_{b}^{\dagger}e^{-i\op H_{I}^{\prime}t/\hbar}\op U_{b}\ket{\psi\left(0\right)} \nonumber \\
	\defeq & \op U_{b}^{\dagger}e^{-i\op H_{I}^{\prime}t/\hbar}\ket{\psi^{\prime}\left(0\right)}.
\end{align}
In terms of the entanglement dynamics, we are not concerned with the basis transformation as it only acts locally on each of the atoms and field modes. If we consider two identical atoms, with frequencies $\omega_{0}$ and dipole strengths 
$\left|d_{j}\right| = d$ coupled to a single EM field mode, then the previous paragraph implies 
that it suffices to consider only a Hamiltonian where both couplings are real, and the atomic 
separation does not enter; thus, there can be no non-trivial distance dependence. However, if we 
consider the two identical atoms coupled to two field modes, then without loss of generality we 
can write the total Hamiltonian of the system as 
\begin{align} 
	\hat{H}= & \hbar \left[ 
	\omega_{0}\left(
	\op{\sigma}_{1}^{\dagger}\op{\sigma}_{1}+\op{\sigma}_{2}^{\dagger}\op{\sigma}_{2}
	\right)
	+\omega_{1}\op 
	a_{1}^{\dagger}\op a_{1}+\omega_{2}\op a_{2}^{\dagger}\op a_{2} 
	\right. \label{eq:TwoModeHaq} \\
	+ & \left. \left( 
	g_{1} \op{\sigma}_{1}^{+}\op a_{1} + g_{2} \op{\sigma}_{1}^{+}\op a_{2} +	g_{1} \op{\sigma}_{2}^{+}\op a_{1} + 
	g_{2} e^{i\phi} \op{\sigma}_{2}^{+}\op a_{2}+ H.C. 
	\right)
	\right],\nonumber 
\end{align}
 where all distance dependence arises from $\phi=\left(k_{2}-k_{1}\right)\left(w_{2}-w_{1}\right)$, and ``H.C.'' denotes the Hermitian conjugate terms. For a pair of counter-propagating modes, we have $k_2 = -k_1 \defeq 2\pi/\lambda$ and hence the parameter $\phi= 4\pi\Delta w/\lambda$.
For simplicity, we will
further assume that $\omega_{1}=\omega_{2}=\omega_{0}$, which implies $g_{1}=g_{2}\defeq g$. 

Our aim
is to get a sense of the variety of different entanglement dynamics
that can result from different separations and initial conditions. To that end, we will first consider
two special cases, which are arguably extreme cases of the general
model, $\phi \mod 2\pi=0$ (where $\Delta w = n \lambda/2$) which we will call the case of symmetric coupling,  
and $\phi \mod 2\pi = \pi$ (where $\Delta w = \left( 1/4 + n/2 \right) \lambda$), which we will call the case of antisymmetric coupling (of the second atom to the second mode).  
Next we will discuss how the qualitative features of entanglement dynamics can differ significantly between these two special cases of atomic separation.  

\section{Quantifying Entanglement}

In order to quantitatively study the entanglement dynamics, we must specify a way of quantifying 
entanglement. For two qubits one may compute the entanglement measure known as the entanglement 
of formation efficiently in terms of the Wootters concurrence \cite{Wooters1998}. In order to 
compute the concurrence, one must choose an arbitrary basis and construct the spin-flipped 
operator \begin{equation} 
\tilde{\rho}\defeq\left(\hat{\sigma}_{y}\otimes\hat{\sigma}_{y}\right)\hat{\rho}^{*}\left(\hat{\sigma}_{y}\otimes\hat{\sigma}_{y}\right),\end{equation}
 where the star denotes the complex conjugation in that basis and
$\hat{\sigma}_{y}$ is the Pauli matrix in the same basis. The concurrence
is \begin{equation}
C\left(\op{\rho}\right)=\max\left\{ 0,\sqrt{\lambda_{1}}-\sqrt{\lambda_{2}}-\sqrt{\lambda_{3}}-\sqrt{\lambda_{4}}\right\} ,\end{equation}
 where the $\lambda_{j}$s are the eigenvalues of the matrix $\op{\rho}\tilde{\rho}$
in decreasing order. The entanglement of formation can then be computed
as \begin{equation}
\mathcal{E}_{F}\left(C\right)=h\left(\frac{1+\sqrt{1-C^{2}}}{2}\right),\end{equation}
 where \begin{equation}
h\left(x\right)=-x\log_{2}\left(x\right)-\left(1-x\right)\log_{2}\left(1-x\right).\end{equation}
 Alternatively, to quantify entanglement one can also use the negativity \cite{VidalWerner2002} (which quantifies the degree to which the state violates the Peres-Horodecki positive partial transpose condition), defined
\begin{equation}
\mathcal{N}\left(\op{\rho}\right)=\frac{\left\Vert \hat{\rho}^{T_{B}}\right\Vert _{1}-1}{2},
\end{equation}
where $\hat{\rho}^{T_{B}}$ is the partial transpose of the density
matrix, defined by \begin{equation}
\bra{j,k}\hat{\rho}^{T_{B}}\ket{l,q}\defeq\bra{j,q}\hat{\rho}\ket{l,k}\end{equation}
 in an arbitrary basis B, and $\left\Vert \cdot\right\Vert _{1}$ is the trace norm. Either of 
these approaches can be used to quantify entanglement.  The negativity is arguably less precise, 
because in some systems it can be zero even for an entangled state, but we will not consider such 
situation.  The advantage of the negativity is that it can be efficiently computed even in a 
large Hilbert space.  We will most often use the concurrence in this article; however, we do 
quantify the entanglement of field states using the negativity.

\section{Special cases\label{sec:Sp}}
As a first step toward evaluating the influence of interatomic separation on entanglement dynamics, 
we now examine two particularly simple cases that one might expect to represent two distinct extremes.
Since we have assumed the two field modes have the same frequency,
rather than using the original modes $F_{1}$ and $F_{2}$ one could
equally well choose a different mode decomposition of the field where
the two modes are replaced by linear combinations $TF_{1}$ and $TF_{2}$
with annihilation operators \begin{align}
\op A_{1} & =\frac{1}{\sqrt{2}}\left(\op a_{1}+\hat{a}_{2}\right) \label{eq:TransMode1} \\
\op A_{2} & =\frac{1}{\sqrt{2}}\left(\op a_{1}-\hat{a}_{2}\right).\label{eq:TransMode2}
\end{align}
These will be two orthogonal standing wave modes.  The assumed translation invariance of the mode 
structure in the direction of the coordinate $w$ allows us to choose the phase of the standing wave 
modes arbitrarily (subject to the constraint of orthogonality), and the mathematical definitions we 
have given here specify that $TF_{1}$ will have an anti-node at the position of atom 1 and $TF_{2}$ 
will have a node at that location.  The coupling to atom 2 then depends on the separation.  
With this mode transformation we can now turn to the special cases of interest.

When $\phi=0$ we have the case of two modes with symmetrical coupling (SC) to the second atom.  
The interaction Hamiltonian can then be written \begin{equation}
\op H_{I}=\sqrt{2}\hbar g\left[\op{\sigma}_{1}^{+}\op A_{1}+\op{\sigma}_{1}^{-}\op A_{1}^{\dagger}+\op{\sigma}_{2}^{+}\op A_{1}+\op{\sigma}_{2}^{-}\op A_{1}^{\dagger}\right]\label{eq:TMSCTransHaq}\end{equation}
 in terms of these transformed modes, so that the atoms only couple
to $\hat{A}_{1}$ and not $\hat{A}_{2}$. This reflects the fact that 
the atoms are separated by $\Delta w = n \lambda/2$, so the second atom 
also lies at a node of $TF_{2}$ and anti-node of $TF_{1}$.
Thus, the Hamiltonian
is equivalent to the model of a single mode symmetrically coupled
to two atoms, a Tavis-Cummings (TC) model, and all work done on that model can be applied here.
 The evolution of the reduced density
matrix of the atom $\op{\rho}_{A}\left(t\right)$ in the SC model
is be the same as in the TC model with a properly transformed
initial state. Namely, if the total system (atoms and modes) is in
an initial state described by the density matrix $\op{\chi}_{SC}\left(0\right)$,
then the appropriate initial density matrix for the TC model is obtained
by making the mode transformation of Eqs. \eqref{eq:TransMode1} and
\eqref{eq:TransMode2} and tracing out the second transformed mode
$TF_{2}$. As this mapping of initial states involves
a partial trace, it is a many-to-one mapping from the SC problem
to the TC problem (and this mapping does not preserve purity). 

In order to
solve the dynamics in the TC model, and by extension the SC model,
we simply compute the time evolution operator expressed in the atomic basis $\{\ket{ee},\ket{eg},\ket{ge},\ket{gg}\}$ directly by exponentiation
(as in, e.g. \cite{KimKnight2003}):
\begin{align}
	\hat{H}_{I}= & \sqrt{2}g\left(\begin{array}{cccc}
	0 & \hat{A}_{1} & \hat{A}_{1} & 0\\
	\hat{A}^{\dagger}_1 & 0 & 0 & \hat{A}_1\\
	\hat{A}^{\dagger}_1 & 0 & 0 & \hat{A}_1\\
	0 & \hat{A}^{\dagger}_1 & \hat{A}^{\dagger}_1 & 0\end{array}\right) \nonumber \\
	\Rightarrow & \, \hat{U}=e^{-i\hat{H}_{I}t}=\left(\begin{array}{cccc}
	\hat{C}_{1} & -i\hat{S}_{1} & -i\hat{S}_{1} & \hat{C}_{2}\\
	-i\hat{S}_{2} & \hat{C}_{3} & \hat{C}_{4} & -i\hat{S}_{3}\\
	-i\hat{S}_{2} & \hat{C}_{4} & \hat{C}_{3} & -i\hat{S}_{3}\\
	\hat{C}_{5} & -i\hat{S}_{4} & -i\hat{S}_{4} & \hat{C}_{6}\end{array}\right)
\end{align}
 where 
\begin{align}
	&
	\begin{aligned}
		\hat{S}_{1} & =\hat{A}_{1}\frac{\sin\left(\sqrt{4\hat{\mathcal{A}}}gt\right)}{\sqrt{2\hat{\mathcal{A}}}} & \qquad
		\hat{S}_{2} & =\frac{\sin\left(\sqrt{4\hat{\mathcal{A}}}gt\right)}{\sqrt{2\hat{\mathcal{A}}}}\hat{A}_{1}^{\dagger} \nonumber \\
		\hat{S}_{3} & =\frac{\sin\left(\sqrt{4\hat{\mathcal{A}}}gt\right)}{\sqrt{2\hat{\mathcal{A}}}}\hat{A}_{1} & \,
		\hat{S}_{4} & =\hat{A}_{1}^{\dagger}\frac{\sin\left(\sqrt{4\hat{\mathcal{A}}}gt\right)}{\sqrt{2\hat{\mathcal{A}}}} \nonumber
	\end{aligned} \\
	& \hat{C}_{1} =1-\hat{A}_{1}\frac{1}{\hat{\mathcal{A}}}\hat{A}_{1}^{\dagger}+\hat{A}_{1} \frac{\cos\left(\sqrt{4\hat{\mathcal{A}}}gt\right)}{\hat{\mathcal{A}}}\hat{A}_{1}^{\dagger} \nonumber \\
	& \hat{C}_{2} =-\hat{A}_{1}\frac{1}{\hat{\mathcal{A}}}\hat{A}_{1}+\hat{A}_{1}\frac{\cos\left(\sqrt{4\hat{\mathcal{A}}}gt\right)}{\hat{\mathcal{A}}}\hat{A}_{1} & & \\
	&
	\begin{aligned}
		\hat{C}_{3} & =\frac{1}{2}\left(\cos\left(\sqrt{4\hat{\mathcal{A}}}gt\right)+1\right) &
		\hat{C}_{4} & =\frac{1}{2}\left(\cos\left(\sqrt{4\hat{\mathcal{A}}}gt\right)-1\right) \nonumber
	\end{aligned} \nonumber \\
	& \hat{C}_{5} =-\hat{A}_{1}^{\dagger}\frac{1}{\hat{\mathcal{A}}}\hat{A}_{1}^{\dagger}+\hat{A}_{1}^{\dagger}\frac{\cos\left(\sqrt{4\hat{\mathcal{A}}}gt\right)}{\hat{\mathcal{A}}}\hat{A}_{1}^{\dagger} \nonumber \\
	& \hat{C}_{6} =1-\hat{A}_{1}^{\dagger}\frac{1}{\hat{\mathcal{A}}}\hat{A}_{1}+\hat{A}_{1}^{\dagger}\frac{\cos\left(\sqrt{4\hat{\mathcal{A}}}gt\right)}{\hat{\mathcal{A}}}\hat{A}_{1} \nonumber \\
	& \hat{\mathcal{A}} \equiv\hat{A}_{1}\hat{A}_{1}^{\dagger}+\hat{A}_{1}^{\dagger}\hat{A}_{1} \nonumber
\end{align}
We observe that there is generally oscillatory behavior for the state dynamics coming from the periodic time dependence and the frequency of these oscillations depends on the number of excitations in the initial field state. Also, a Fock state has a well defined frequency.

When $\phi=\pi$ we have a two-mode model with asymmetric coupling
(AC) to the second atom.  In terms of the transformed modes of Eqs.~\eqref{eq:TransMode1}
and \eqref{eq:TransMode2} the Hamiltonian can be written as \begin{equation}
\op H_{I}=\sqrt{2}\hbar g\left[\op{\sigma}_{1}^{+}\op A_{1}+\op{\sigma}_{1}^{-}\op A_{1}^{\dagger}+\op{\sigma}_{2}^{+}\op A_{2}+\op{\sigma}_{2}^{-}\op A_{2}^{\dagger}\right].\label{eq:TMACTransHaq}\end{equation}
 In this case, rather than both atoms coupling to one mode we see
that Atom 1 couples only to transformed mode $TF_{1}$ while Atom
2 couples only to $TF_{2}$.  Because in this case $\Delta w = \left( 1/4 + n/2 \right) \lambda$,
atom 2 now lies at a node of $TF_{1}$ and anti-node of $TF_{2}$.
This situation is equivalent
to a model comprised of two subsystems that are totally isolated from
one another, each composed of a single atom coupled to a single mode.
We will call this the double Jaynes-Cummings (DJC) model. This sort
of model with isolated subsystems is common to the study of entanglement
sudden death \cite{YuEberly2004,YuEberly2009}, and the DJC model
specifically has been studied \cite{Chan2009,YonacYuEberly2006,YonacYuEberly2007}, 
so the mapping allows those investigations to bear on this model.

As in the previous case, the evolution of $\op{\rho}_{A}\left(t\right)$
for the AC model should be the same as given by the DJC model with
the proper mapping of initial states. This implies that there can, for example, be no dynamical
generation of atomic entanglement in the AC model unless the DJC initial
field state obtained by the mapping is entangled. In the DJC model we can write the
unitary time-evolution operators for the two separate non-interacting atom-field
subsystems as $\hat{U}_{1}$ and $\hat{U}_{2}$, and then the total
time evolution operator is $\op U=\op U_{1}\otimes\op U_{2}$. We
again compute the two subsystem unitary time evolution operators by direct exponentiation
to obtain 
\begin{align}
\hat{U}_{1}=e^{-i\hat{H}_{1}t}= & \left(\begin{array}{cccc}
\hat{C}_{11} & 0 & -i\hat{S}_{11} & 0\\
0 & \hat{C}_{11} & 0 & -i\hat{S}_{11}\\
-i\hat{S}_{12} & 0 & \hat{C}_{12} & 0\\
0 & -i\hat{S}_{12} & 0 & \hat{C}_{12}\end{array}\right)\\
\hat{U}_{2}=e^{-i\hat{H}_{2}t}= & \left(\begin{array}{cccc}
\hat{C}_{21} & -i\hat{S}_{21} & 0 & 0\\
-i\hat{S}_{22} & \hat{C}_{21} & 0 & 0\\
0 & 0 & \hat{C}_{22} & -i\hat{S}_{21}\\
0 & 0 & -i\hat{S}_{22} & \hat{C}_{22}\end{array}\right)\end{align}
 with \begin{align}
\hat{C}_{i1} & =\cos(\sqrt{2\hat{A}_{i}\hat{A}_{i}^{\dagger}}gt) & 
\hat{C}_{i2} & =\cos(\sqrt{2\hat{A}_{i}^{\dagger}\hat{A}_{i}}gt) \\
\hat{S}_{i1} & =\frac{\sin(\sqrt{2\hat{A}_{i}\hat{A}_{i}^{\dagger}}gt)}{\sqrt{\hat{A}_{i}\hat{A}_{i}^{\dagger}}}\hat{A}_{i} & 
\hat{S}_{i2} & =\hat{A}_{i}^{\dagger}\frac{\sin(\sqrt{2\hat{A}_{i}^{\dagger}\hat{A}_{i}}gt)}{\sqrt{\hat{A}_{i}^{\dagger}\hat{A}_{i}}}.\nonumber 
\end{align}

In order to illustrate the variety of behavior that can arise among
these models, we examine
the entanglement dynamics of a selection of initial states in which
the atoms are uncorrelated with the fields. In each case we select the
atomic state from the set of pure states $\left\{ \ket{gg},\ket{ee},\ket{eg},\ket{\Phi},\ket{\Psi}\right\} $,
where $\ket{\Phi}\defeq\left(\ket{ee}+\ket{gg}\right)/\sqrt{2}$,
$\ket{\Psi}\defeq\left(\ket{eg}+\ket{ge}\right)/\sqrt{2}$.
The initial field state will be either a product
of Fock states $\ket{n_N,m_{N}}$, a product of Glauber coherent
states $\ket{\alpha_{c},\beta_{c}}$, a product of squeezed vacuum
states $\ket{\xi_{sq},-\xi_{sq}}$ , a two-mode squeezed vacuum state
(TMSS) $\ket{\xi,0,0}$ , a thermal state $\op{\rho}^{(1)}_{th}\otimes\op{\rho}^{(2)}_{th}$ (with
both modes having equal temperature), the pure state $\ket{\eta_{nm}}$,
or the mixed state $\op{\rho}_{nm}$. The two-mode squeezed state is defined as the state resulting from the action of the two-mode squeezing operator $\hat{S}(\xi)=e^{\left(\xi^{\ast}\hat{a}_1\hat{a}_2-\xi\hat{a}_1^{\dagger}\hat{a}_2^{\dagger}\right)}$ on vacuum. The state $\ket{\eta_{nm}}$ is
the result of mapping the state $\ket{n_{N},m_{N}}$ in the original
modes of the AC problem to the transformed modes equivalent to
the DJC problem, with 
\begin{align}
\ket{\eta_{nm}}=&\frac{1}{\sqrt{2^{m+n}m!n!}}\sum_{k=0}^{n}\sum_{l=0}^{m}\leftexp nC_{k}\leftexp mC_{l}\sqrt{(m+n-k-l)!}\\
&\sqrt{(k+l)!}(-1)^{l}\ket{m+n-k-l}\ket{k+l},\nonumber
\end{align}
and $\hat{\rho}_{nm}\defeq\tr_{TF_{2}}\left[\ket{\eta_{nm}}\bra{\eta_{nm}}\right]$ is the state in the TC model that gives equivalent evolution to the state $\ket{n_{N},m_{N}}$ in the SC model. To find the atom-atom density matrix we 
trace out the field modes from the time evolved state of the closed atom-field system.


\begin{table*}[bht]
\begin{tabular}{|l|l|c|c|c|c|c|}
\hline 
\multicolumn{2}{|c}{Initial field State} & \multicolumn{5}{|c|}{Atomic State}\tabularnewline
\hline 
SC  & TC  & A. $\ket{ee}$  & B. $\ket{eg}$  & C. $\ket{gg}$  & D. $\ket{\Phi}$  & E. $\ket{\Psi}$\tabularnewline
\hline
\hline 
1. $\ket{n_{N},m_{N}}$  & $\hat{\rho}_{nm}$  & No  & Yes, DI  & Yes\footnote[1]{No entanglement for $n_N=m_N$}, DI/SD  & SD  & SD\tabularnewline
\hline 
2. $\ket{\eta_{nm}}$ & $\ket{n_{N}}$  & No  & Yes, DI  & Yes, SD  & SD  & SD\tabularnewline
\hline 
3. $\hat{\rho}_{th}$  & $\hat{\rho}_{th}$  & No  & Yes, DI  & Yes, SD  & AL/SD & SD \tabularnewline
\hline 
4. $\ket{\frac{1}{\sqrt{2}}\left(\alpha_{c}+\beta_{c}\right),\frac{1}{\sqrt{2}}\left(\alpha_{c}-\beta_{c}\right)}$  & $\ket{\alpha_{c}}$  & Yes, AL/SD  & Yes, SD  & Yes, AL/SD  & AL/SD  & AL/SD \tabularnewline
\hline 
5. $\ket{\xi_{sq},-\xi_{sq}}$  & $\hat{\rho}_{th}$  & No  & Yes, DI  & Yes, SD  & AL/SD & SD \tabularnewline
\hline 
6. $\ket{\xi,0,0}$  & $\ket{\xi_{sq}}$  & Yes, AL/SD  & Yes,SD  & Yes, AL/SD  & AL  & SD \tabularnewline
\hline
\multicolumn{1}{c}{} & \multicolumn{1}{c}{} & \multicolumn{1}{c}{} & \multicolumn{1}{c}{} & \multicolumn{1}{c}{} & \multicolumn{1}{c}{} & \multicolumn{1}{c}{}\tabularnewline
\hline 
\multicolumn{2}{|c}{Initial field State} & \multicolumn{5}{|c|}{Atomic State}\tabularnewline
\hline 
AC  & DJC  & A. $\ket{ee}$  & B. $\ket{eg}$  & C. $\ket{gg}$  & D. $\ket{\Phi}$  & E. $\ket{\Psi}$\tabularnewline
\hline
\hline 
1. $\ket{n_{N},m_{N}}$  & $\ket{\eta_{nm}}$  & Yes\footnotemark[1], SD  & Yes\footnotemark[1], SD  & Yes\footnotemark[1], SD/DI  & SD  & SD/AL \tabularnewline
\hline 
2. $\ket{\eta_{nm}}$  & $\ket{n_{N},m_{N}}$  & No  & No  & No  & SD  & SD \tabularnewline
\hline 
3. $\hat{\rho}_{th}$  & $\hat{\rho}_{th}$  & No  & No  & No  & SD  & SD \tabularnewline
\hline 
4. $\ket{\alpha_{c},\beta_{c}}$  & $\ket{\frac{1}{\sqrt{2}}\left(\alpha_{c}+\beta_{c}\right),\frac{1}{\sqrt{2}}\left(\alpha_{c}-\beta_{c}\right)}$  & No  & No  & No  & SD  & SD\tabularnewline
\hline 
5. $\ket{\xi_{sq},-\xi_{sq}}$  & $\ket{\xi,0,0}$  & Yes, SD  & Yes, SD  & Yes, SD  & SD  & SD \tabularnewline
\hline 
6. $\ket{\xi,0,0}$  & $\ket{\xi_{sq},-\xi_{sq}}$  & No  & No  & No  & SD  & SD \tabularnewline
\hline
\end{tabular}

\caption{Entanglement dynamics for two modes symmetrical coupling  (SC) with $\phi=0$ and anti-symmetrical coupling with $\phi=\pi$. Columns A-C list whether there is an entanglement generation in an initially separable atomic state (yes or no). The dynamical phenomena observed in columns A-E are listed as entanglement sudden death (SD), entanglement dies for only an instant (DI), entanglement remains non-zero at all times and so is ``always living'' (AL). A `/' denotes that both kinds of dynamics are present depending on the particular initial state chosen from within the class indicated.}
\label{phi0}
\end{table*}

We summarize
our findings for the entanglement behavior given the various initial
states considered in the four models in Table \ref{phi0}, listing
the equivalent SC-TC cases and
 AC-DJC cases.
When discussing entanglement sudden death, we adopt the usage of the term
as in \cite{YonacYuEberly2006} in applying it only to instances where
the entanglement goes to zero for some finite (non-zero) duration. In the case where entanglement goes to zero only for an instant during the time evolution we refer to it as death for an instant (DI). If there is a non-zero entanglement at all times once it is generated in the system then we label it as being ``always living'' (AL).  In the context of that general summary, we now turn to commenting on some specific features.

\subsection{Entanglement generation in the AC model --- Transfer of correlations}

By reducing the SC and AC models to the TC and DJC models, respectively, we see that the change in atomic separation leads to 
very different entanglement dynamics.  In the case of the AC model, we have a mapping to the dynamics of the DJC model, where each atom interacts only with a separate field mode.  So in the DJC model, the dynamics cannot increase entanglement between the two atom-field subsystems; therefore, if the atomic state is not entangled initially then it will remain separable, unless there is an initial entanglement between the field modes that can be transferred to the atoms by the dynamics.  Knowing this, we can see that any initial field state for the AC model that maps to a separable DJC field state fails to generate entanglement.  The nature of the mapping means that even some entangled field states will fail to generate atomic entanglement in the AC model, while some separable states will map to an entangled DJC state and will generate entanglement.

\begin{figure*}[ht]
\subfloat[$\ket{ee}$]{\label{djctmssee}\includegraphics[width=0.5 \textwidth]{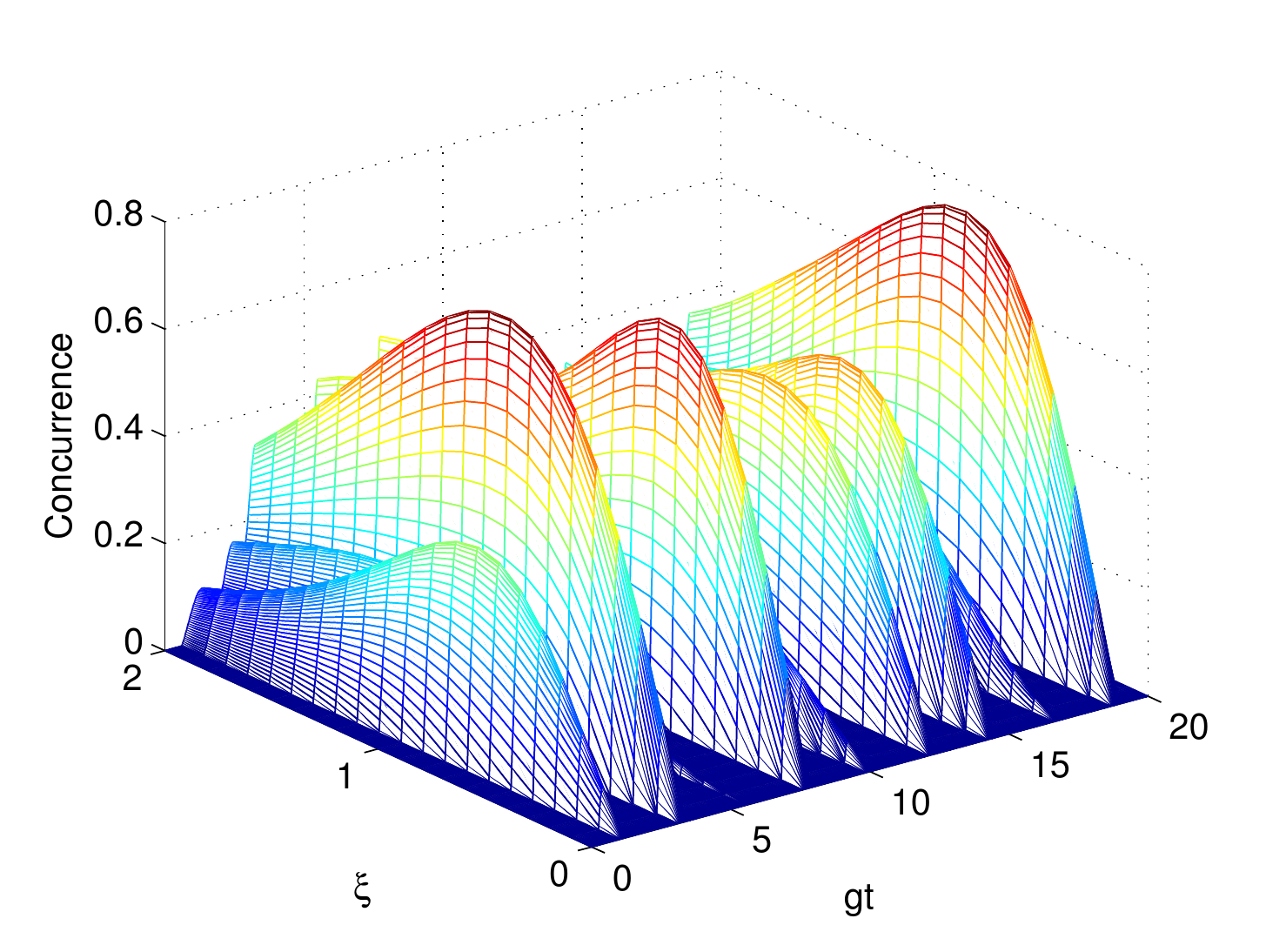}}
\subfloat[$\ket{eg}$ ]{\label{djectmsseg}\includegraphics[width=0.5 \textwidth]{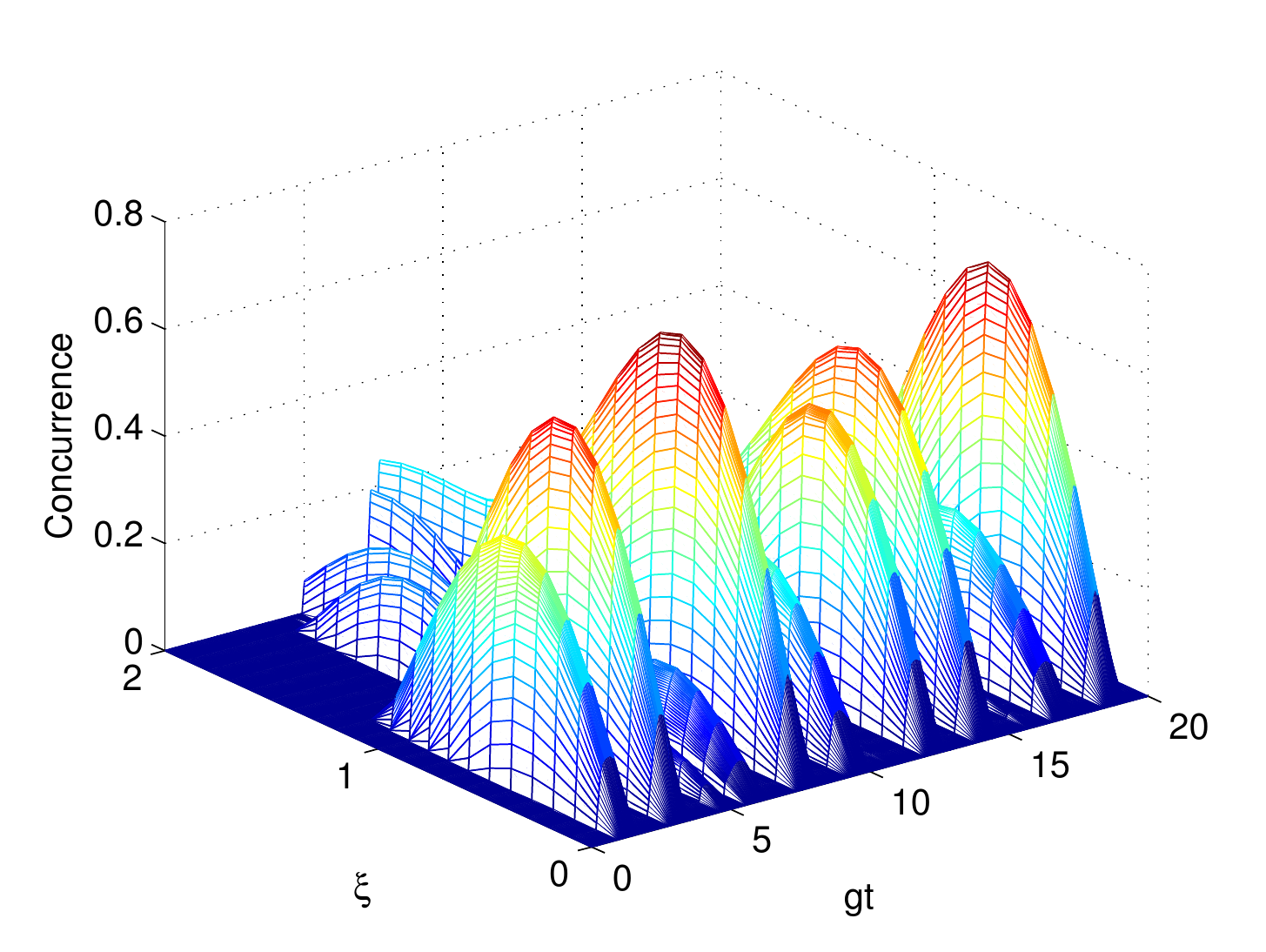}}
\caption{Entanglement dynamics for a two-mode squeezed state $(\ket{\xi,0,0})$ in DJC or product of single mode squeezed states $(\ket{\xi_s,-\xi_s,0})$ in AC interacting with different initial atomic states. Entanglement generation from transfer of correlations from the field, optimal squeezing for maximum transfer of correlations can be seen.}
\label{djctmss}
\end{figure*}

For example, when considering the AC model with an initial field state that is a product of squeezed states $\ket{\xi_{sq},-\xi_{sq}}$ we find that entanglement is dynamically generated as shown in Fig.~\ref{djctmss}. One can understand this from the fact that the state $\ket{\xi_{sq},-\xi_{sq}}$ maps to the DJC model with an initial TMSS, so the generation of atomic entanglement occurs simply because the dynamics transfers the entanglement between the field modes to the atoms.  Further, we find that there is an optimal squeezing parameter value for entanglement generation when the field state is close to being a maximally entangled qubit state $(\alpha_0(\xi)\ket{00}+\alpha_1(\xi)\ket{11})$, with $\alpha_1(\xi)$ and $\alpha_0(\xi)$ being comparable. 
On increasing the squeezing parameter further there are contributions from higher Fock states which decrease the transfer of entanglement.

\begin{figure}[ht]
\includegraphics[width=0.5 \textwidth]{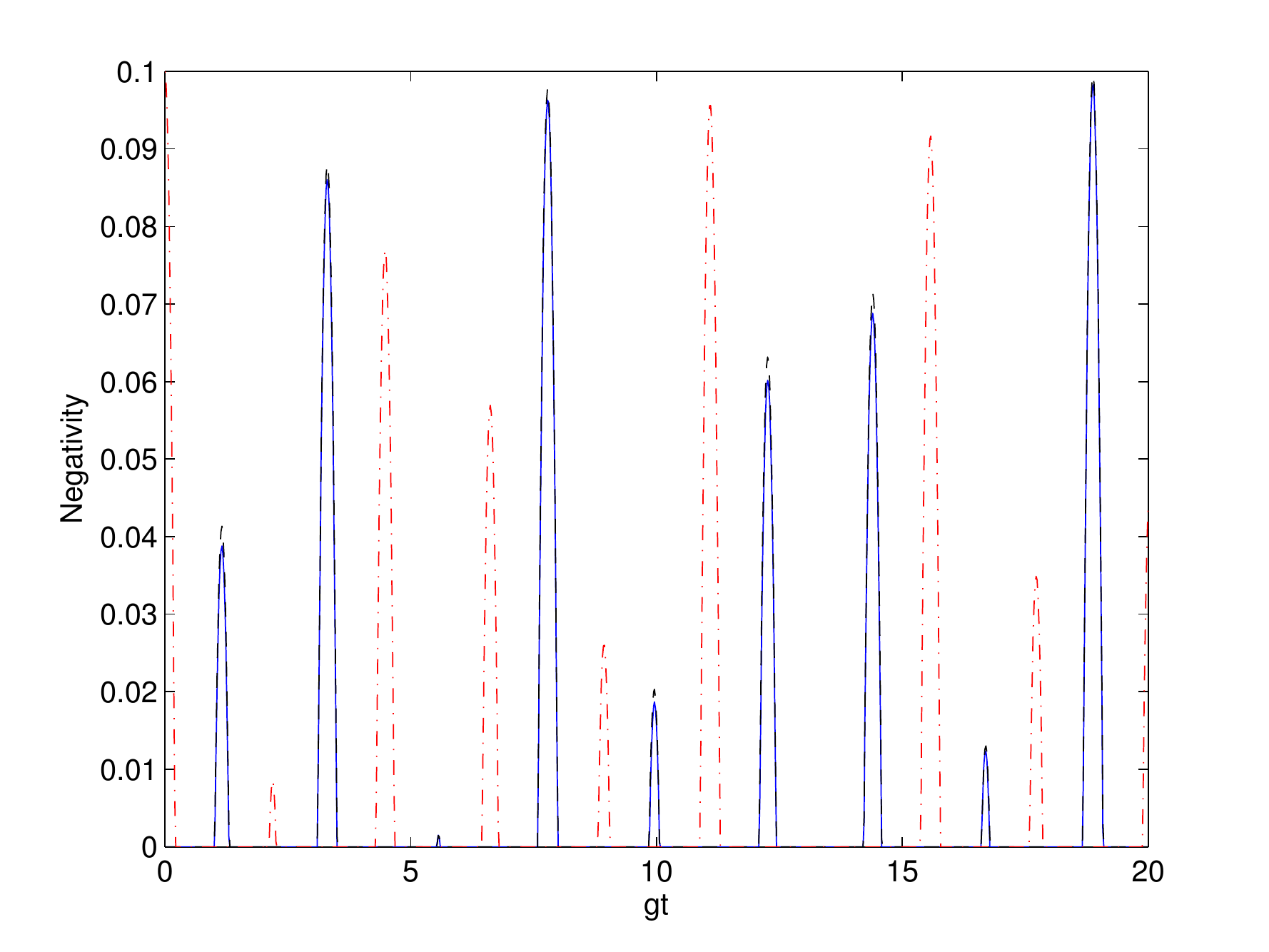}
\caption{Small squeezing entanglement dynamics for a two-mode squeezed state in the DJC model interacting with an initially separable atomic state $\ket{ee}$, entanglement is generated in the atomic subsystem via transfer of field-field correlations. The red dashed line is for the field-field entanglement, blue solid line is the low temperature approximation of the atom-atom entanglement and black dashed is for the exact atom-atom entanglement}
\label{lowtemp}
\end{figure}

Transfer of field-field ($TF_{1}$-$TF_{2}$)
correlations can be explicitly seen in a small squeezing approximation when a two mode squeezed state interacts with an initially separable atomic state $\ket{ee}$.
Since the probability of higher photon numbers is small, by restricting to
the 4-dimensional subspace of lowest energy states in the Fock basis for $TF_{1}$-$TF_{2}$ we obtain the negativities for the two subsystems as,
\begin{align}
\mathcal{N}_{A_{1}-A_{2}}\approx & \vert\min(s_{1}^{2}c_{1}^{2}-\xi s_{1}^{2}c_{2}^{2},0)\vert\\
\mathcal{N}_{TF_{1}-TF_{2}}\approx & \vert\min(s_{1}^{2}c_{1}^{2}-\xi c_{1}^{2}c_{2}^{2},0)\vert.\nonumber 
\end{align}
where $s_1=\sin(\sqrt{2}gt)$, $c_1=\cos(\sqrt{2}gt)$ and $c_2=\cos(2gt)$.  Fig.~\ref{lowtemp} shows how the dynamics of the system transfer the entanglement between field-field subsystem and atom-atom repeatedly.

\subsection{Entanglement generation in the TC model and initial states with identical entanglement dynamics}

Turning to the SC model, the correspondence drawn to the TC model leads to the 
essential feature that the map for the initial field states from SC to TC is many to one; the set of initial states that have 
the same reduced density matrix for the first transformed
mode $TF_{1}$ have identical entanglement dynamics in terms of atom-atom
entanglement. 
As a counterintuitive example of this feature we note 
that a squeezed state of the form $\ket{\xi_{sq},-\xi_{sq}}$ and
a thermal field in the SC model give the same entanglement dynamics provided the
average number of photons in the thermal field corresponds to that
in the squeezed state ($\bar{n}_{th}=\sinh^{2}\left\vert \xi_{sq}\right\vert $).

We can also turn to previous work on the TC model to expose interesting features of the SC model.  As had been shown in \cite{TessierDeutsch2003}, in the TC model an atomic state $\ket{ee}$ interacting with any Fock state $\ket n$ in the field mode never gets entangled.
This can be generalized to any initial field state with a density matrix that is diagonal in the Fock basis as follows:  If one defines 
the time evolved state $\ket{\psi_{n}}\equiv\hat{U}\ket{ee}\ket n$, then the atom-atom density matrix given as $\hat{\rho}^{(n)}=\tr_{F}\left[\ket{\psi_{n}}\bra{\psi_{n}}\right]$ remains separable.
Extending to a general field density matrix diagonal in the Fock basis, the time
evolved atom-atom density matrix given as $\hat{\rho}_{12}=\tr_{F}\left[\sum_{n}{P_{n}\hat{\rho}^{(n)}}\right]$
is clearly a convex sum of separable density matrices, and hence there is no entanglement generation.

\begin{figure*}[htb]
\subfloat[$\ket{eg}$ with $\ket{\eta_{nm}}$ in SC or $\ket{n}$ in TC]{\label{egn}\includegraphics[width=0.45 \textwidth]{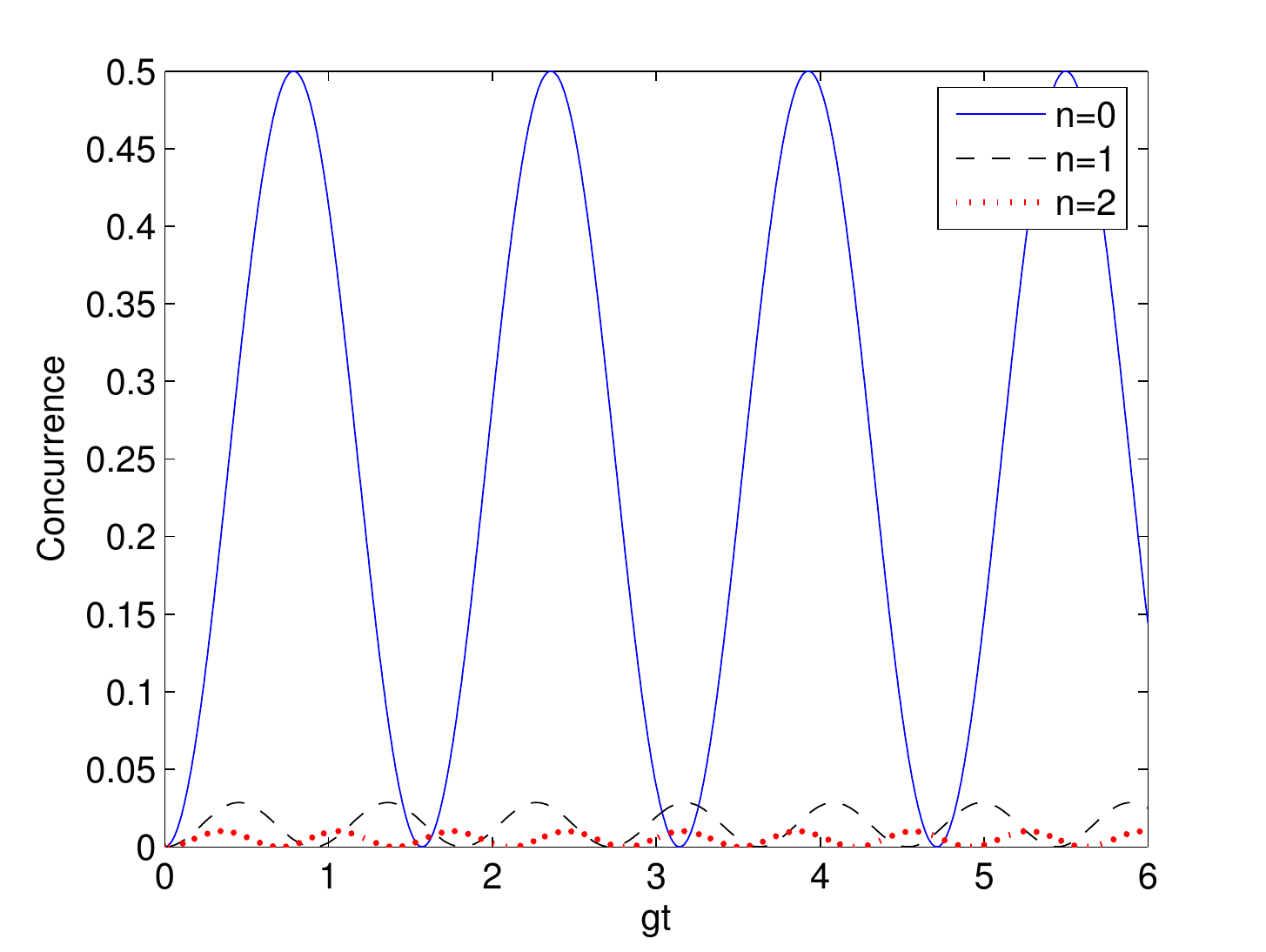}}
\subfloat[$\ket{gg}$ with $\ket{\eta_{nm}}$ in SC or $\ket{n}$ in TC]{\label{ggn}\includegraphics[width=0.45 \textwidth]{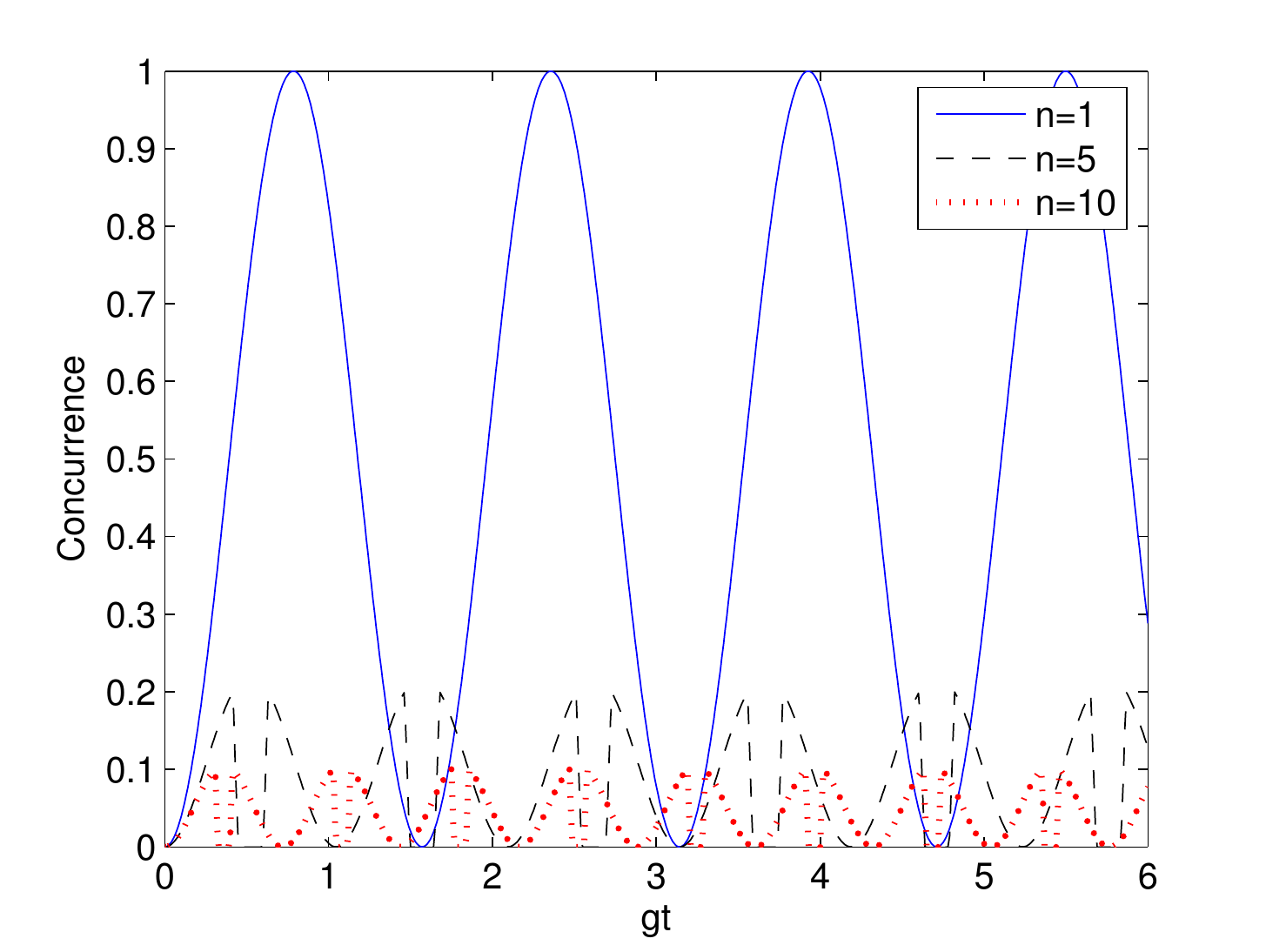}}\\
\subfloat[$\ket{eg}\bra{eg}\otimes\hat{\rho_{th}}$ in TC ]{\label{egth}\includegraphics[width=0.45 \textwidth]{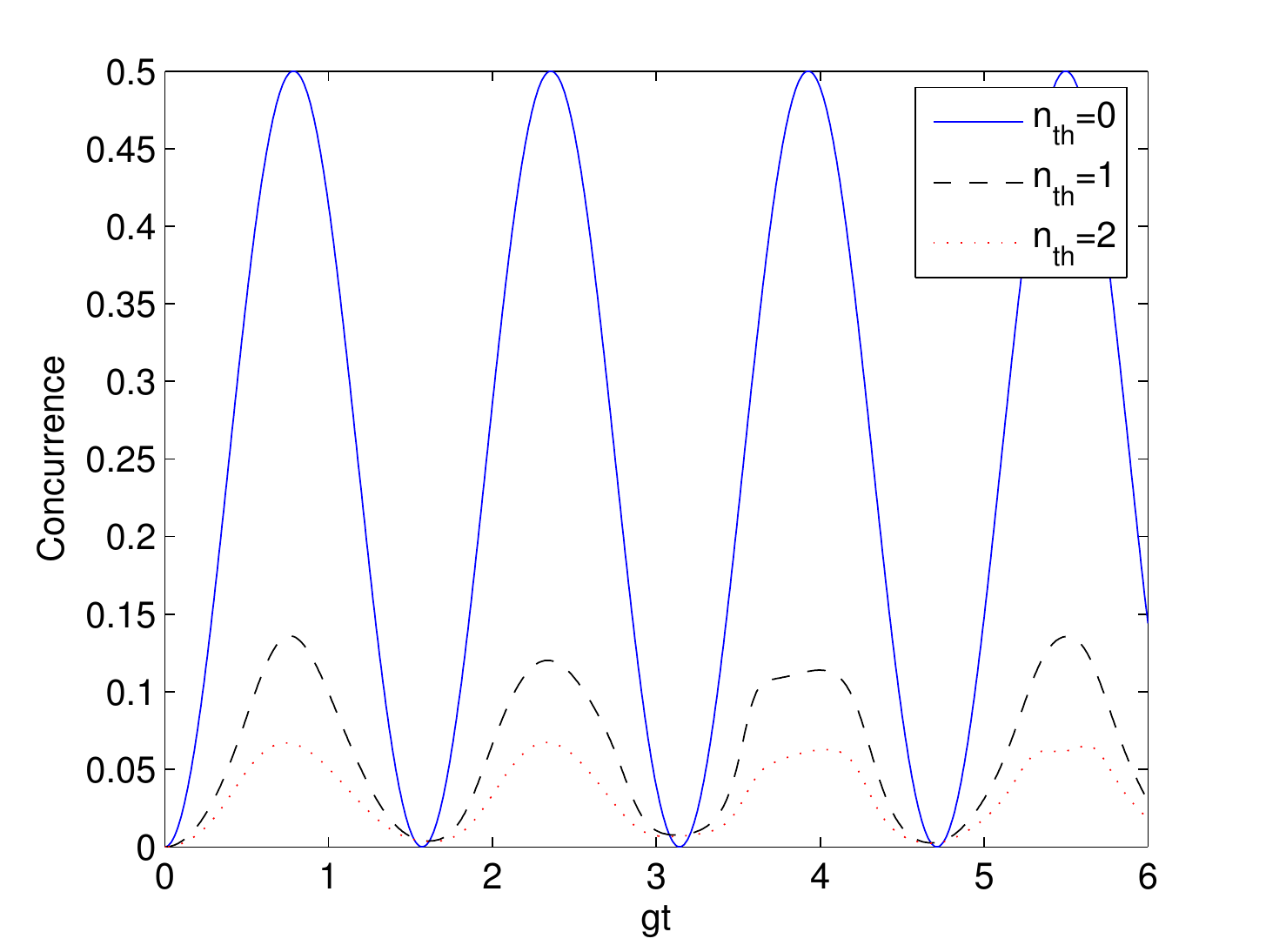}}
\subfloat[$\ket{gg}\bra{gg}\otimes\hat{\rho}_{th}$ in TC ]{\label{ggth}\includegraphics[width=0.45 \textwidth]{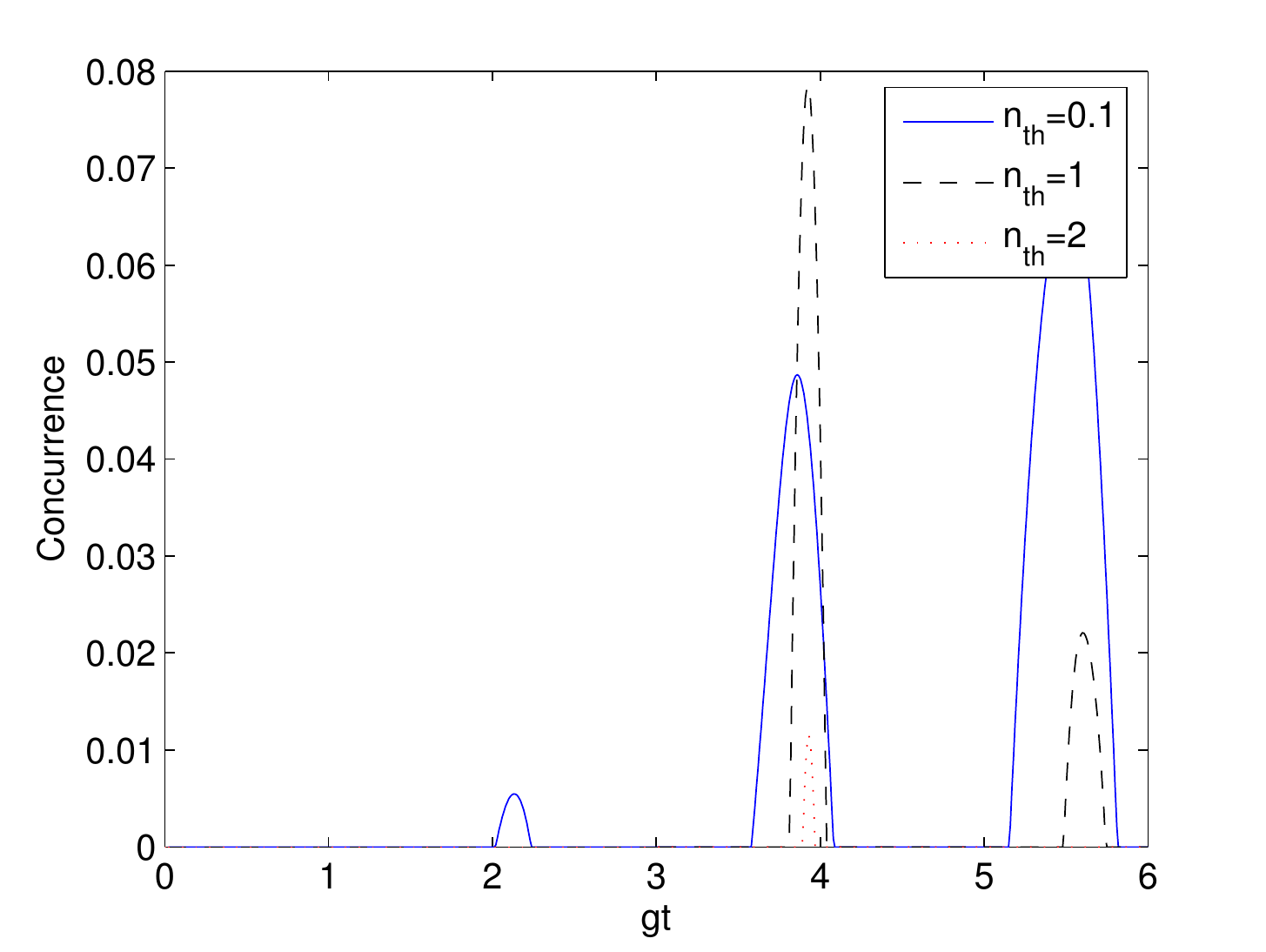}}\\
\subfloat[$\ket{eg}$ with $\ket{n_N,m_N}$ in SC or $\hat{\rho}_{nm}$ in TC  ]{\label{egnm}\includegraphics[width=0.45 \textwidth]{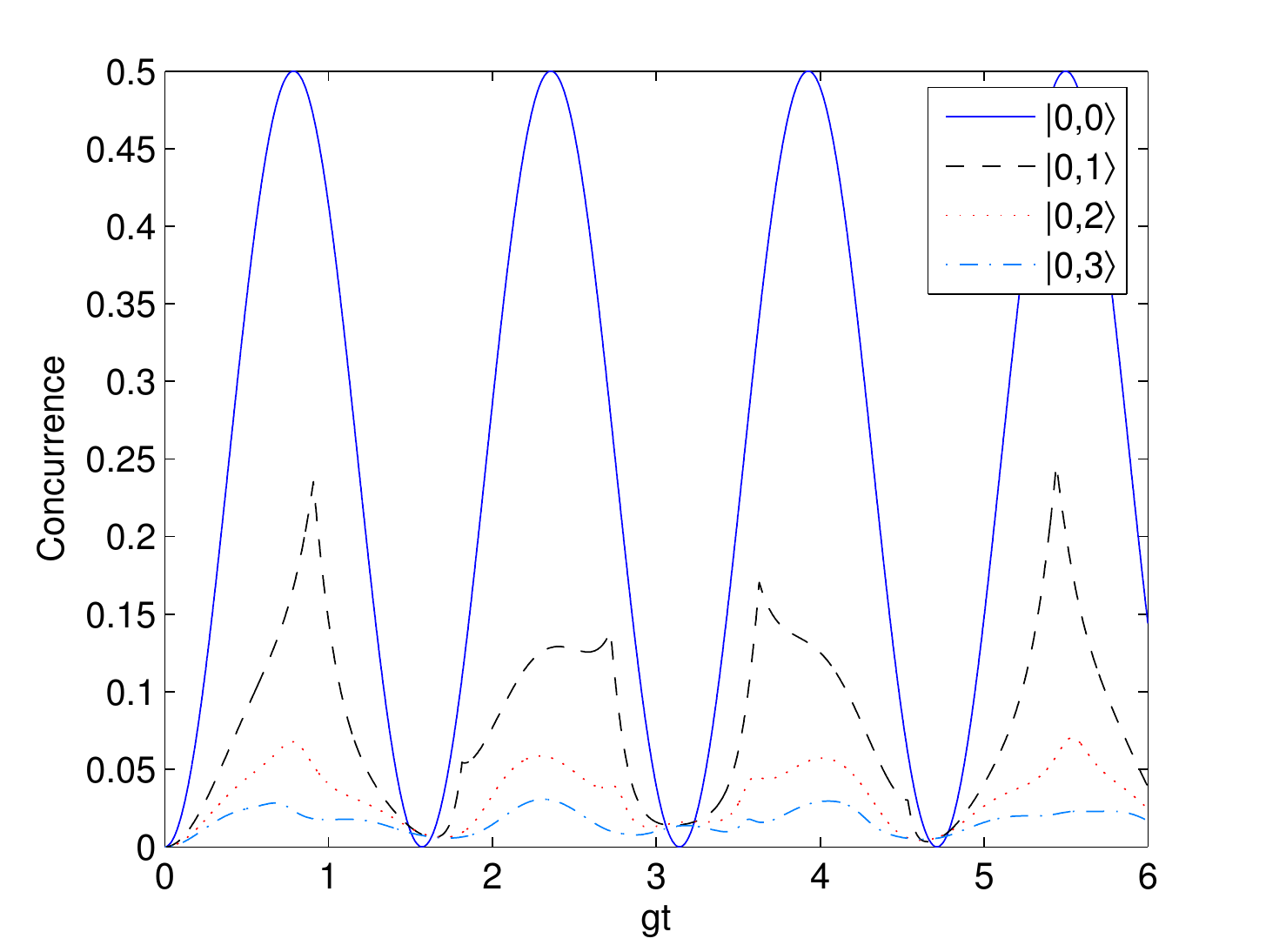}}
\subfloat[$\ket{gg}$ with $\ket{n_N,m_N}$ in SC or $\hat{\rho}_{nm}$ in TC]{\label{ggnm}\includegraphics[width=0.45 \textwidth]{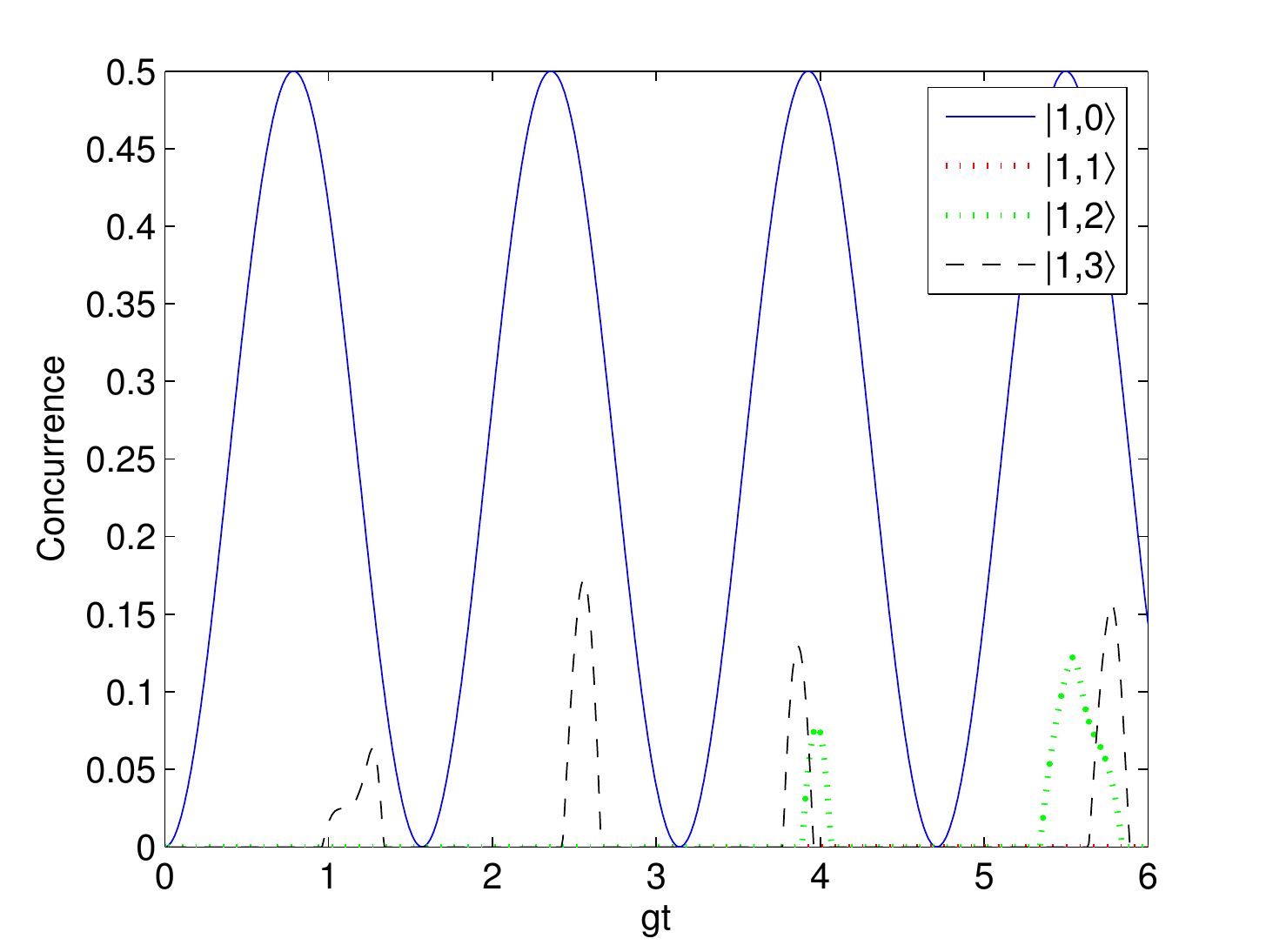}}\\
\caption{Entanglement dynamics in presence of initial field states diagonal in the Fock basis of the TC - no entanglement is seen for initial atomic state $\ket{ee}$, only DI is observed for the atomic state $\ket{eg}$ and maximum entanglement for $\bar{n}\approx1$ for $\ket{gg}$}
\label{diagonal}
\end{figure*}

Looking at the initial state $\ket{eg}\ket n$ in the TC case, we observe no SD in Fig.~\ref{egn}. This can be explained by considering the state as a superposition $\ket{eg}=\frac{1}{2}\left(\ket{eg}+\ket{ge}\right)+\frac{1}{2}\left(\ket{eg}-\ket{ge}\right)$,
where due to the symmetry of the coupling constants the maximally entangled dark state $\left(\ket{eg}-\ket{ge}\right)/\sqrt{2}$ does
not interact with the field. It is only momentarily during the evolution that the state of the system returns to being the original separable superposition $\ket{eg}$. As a result we always observe some entanglement between the two atoms for an initial field density matrix diagonal in the Fock basis.
The dynamics of the $\ket{gg}\ket n$ state in the TC model, shown in Fig.~\ref{ggn}, exhibit SD in general except for the special case of $n=1$ where because of symmetry reasons the state oscillates between the states $\ket{gg}\ket 1$ and $\ket{\Psi}\ket 0$, going from being separable to maximally entangled. Hence, any density matrix of the field modes diagonal in the Fock basis with a high component of $\ket 1\bra 1$ would generate more entanglement in general.

Apart from the Fock state in the TC model and a thermal field,
another example of having a initial field density matrix diagonal
in the Fock basis is to have a Fock state $\ket{n,m}$ in the SC
which corresponds to the TC density matrix 
\begin{align}
	\hat{\rho}_{nm}\equiv\sum_{k,p=0}^{n}\sum_{l,q=0}^{m}{\frac{\kappa_{mnkl}\ket{m+n-k-l}\bra{m+n-k-l}}{2^{m+n}n!m!}}
\end{align}
 where $\kappa_{mnkl}=\leftexp nC_{k}\leftexp nC_{p}\leftexp mC_{l}\leftexp mC_{q}\delta_{k+l,p+q}(m+n-k-l)!(k+l)!(-1)^{l}$.
For this state again we see no entanglement generation for $\ket{ee}$, DI for $\ket{eg}$ and maximal entanglement
in $\ket{gg}\ket{10}$, as shown in Figs.~\ref{egnm} and \ref{ggnm}.
 
Another point we observe is that for an initial
state $\ket{ee}\ket{n,n}$ there is no entanglement generation, which
is a common feature between SC and AC. A Fock state $\ket{n,m}$
in AC transforms into an entangled state in the DJC model, so we
expect to have entanglement generation in the system for an initially
separable atomic state. As an exception we find that for $n=m$, if
there is no atom-atom entanglement to begin with then the atoms remain
separable evolving to a diagonal density matrix.%
 This is counterintuitive in the sense that there is no entanglement transfer to the atomic subsystem from the entangled field modes. This feature can be explained by considering the initial field state $\ket{\eta_{nn}}=\frac{1}{2^n n!}\sum_{k=0}^n{\leftexp{n}C_k(-1)^k\sqrt{2k!(2n-2k)!}\ket{2n-2k,2k}}$. If the atoms are initially in the state $\ket{ee}$ then time evolution will lead to an entanglement of atomic and field states such that detecting whether the number of photons in each mode is even or odd tells us the state of the two atoms.  On tracing out the field this gives us a diagonal atom-atom
density matrix with no atom-atom entanglement.  The same is true for the initial atomic states $\ket{eg}$ and $\ket{gg}$.

\subsection{Selected contrasts in entanglement dynamics}

\begin{figure*}[ht]
\subfloat[$\ket{ee}$ - No SD once entanglement is generated if the state is sufficiently squeezed]{\label{ssqee}\includegraphics[width=0.5 \textwidth]{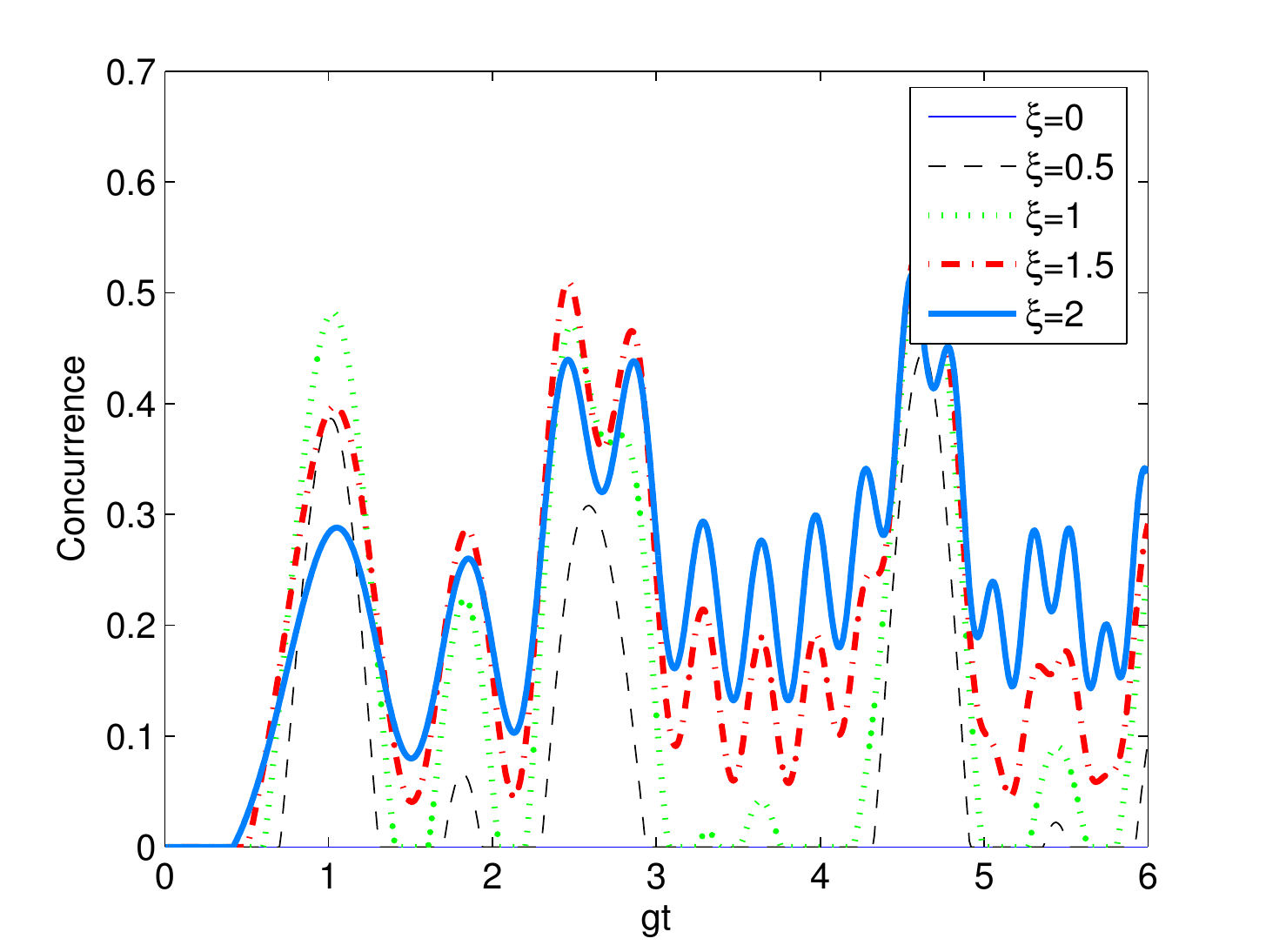}}
\subfloat[$\ket{eg}$ - increased squeezing destroys entanglement in this case]{\label{ssqeg}\includegraphics[width=0.5 \textwidth]{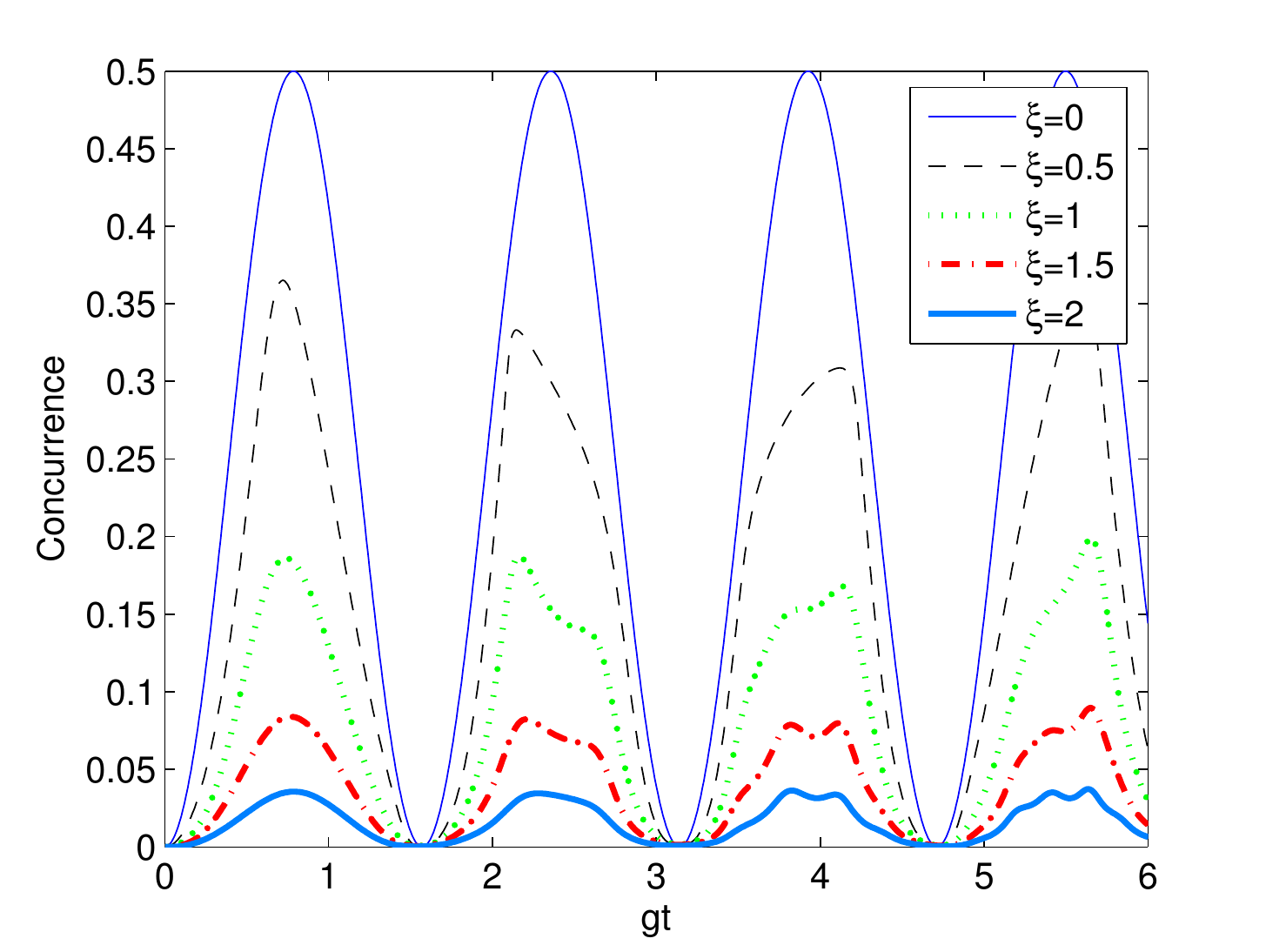}}\\
\subfloat[$\ket{gg}$ - SD disappears on increasing the squeezing]{\label{ssqgg}\includegraphics[width=0.5 \textwidth]{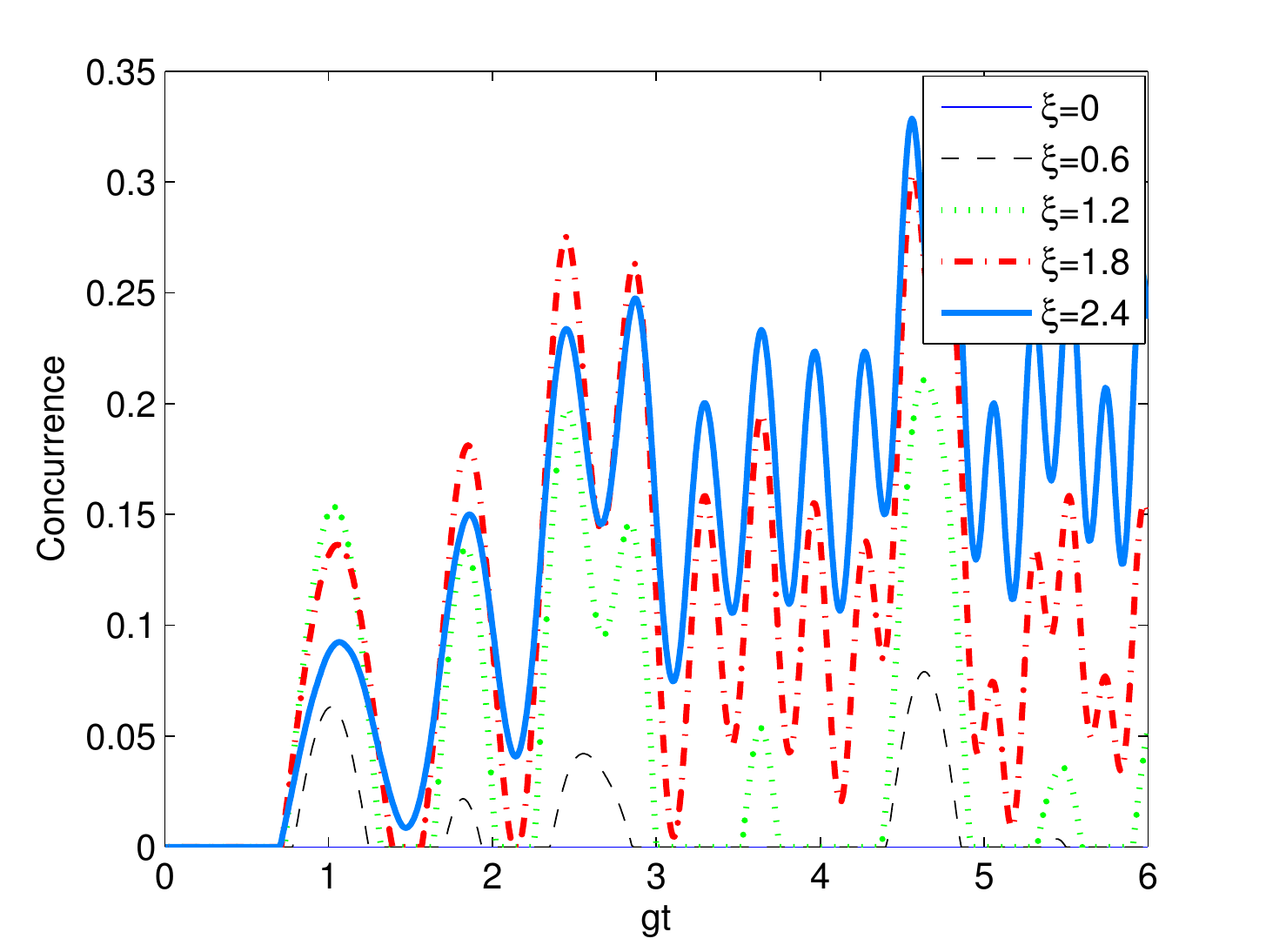}}
\subfloat[$\ket{\Phi}$ - State moves towards being maximally entangled at all times as squeezing is increased]{\label{ssqeegg}\includegraphics[width=0.5 \textwidth]{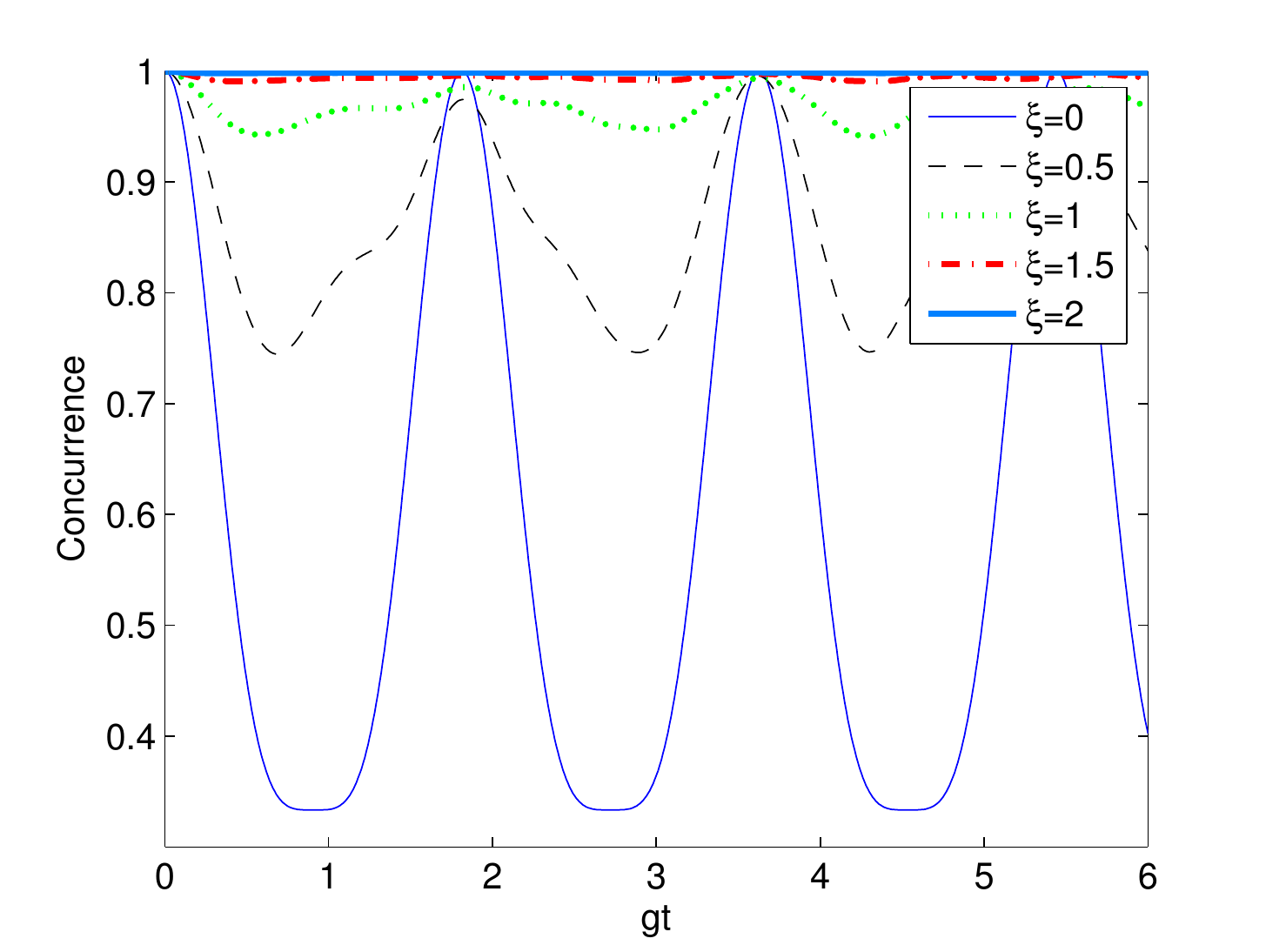}}\\
\caption{Entanglement dynamics for a single mode squeezed state $(\ket{\xi_s})$ in TC or two-mode squeezed state $(\ket{\xi,0,0})$ in SC interacting with different initial atomic states}
\label{ssq}
\end{figure*}

\begin{figure}[ht]
\includegraphics[width=0.5 \textwidth]{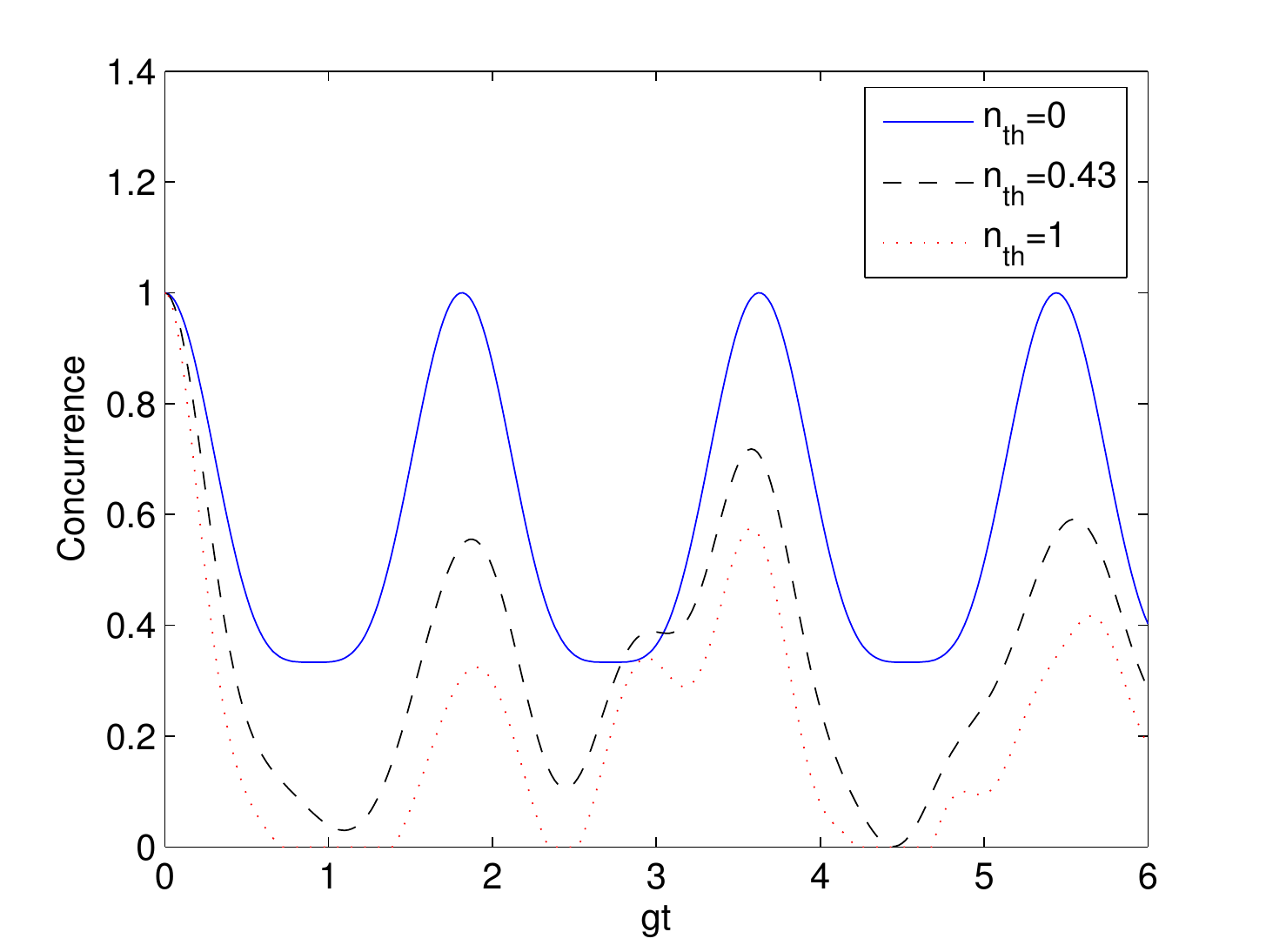}
\caption{Entanglement dynamics for a single mode thermal field interacting with an initially entangled atomic state $\ket{\Phi}$ in the SC model. SD occurs after a certain threshold temperature.}
\label{singlethermeegg}
\end{figure}

While Table \ref{phi0} encapsulates the qualitative features of entanglement dynamics for many states, it may be useful to point to a few interesting examples.  Note that many familiar initial field states that are separable in the DJC model fail to dynamically generate entanglement in the AC model.  In contrast, entanglement generation is a common feature in the SC model for the same selection of field states. For example, starting with a two-mode squeezed state in the field modes does not generate any entanglement in AC, because the corresponding field state in DJC possesses no correlations amongst $TF_{1}$ and $TF_{2}$.
 On the other hand, a two-mode squeezed state in the symmetrically coupled case generates entanglement in the initially separable atomic states as can be seen in Fig.~\ref{ssq}. Furthermore, for the separable initial atomic states $\ket{ee}$ and $\ket{gg}$, if the field is sufficiently squeezed, entanglement is dynamically generated and once generated sustains forever, protected from sudden death.  This behavior is shown in Figs.~\ref{ssqee} and \ref{ssqgg}.  If the atoms are initially in the entangled atomic state $\ket{\Phi}$, then Fig.~\ref{ssqeegg} shows that the entanglement stays protected as the state becomes almost a dark state for large squeezing parameter values.
One can also see entanglement protection for a thermal field in the SC model so long as the thermal average photon number is below a threshold value $\bar{n}_{crit}\approx0.43$ (Fig.~\ref{singlethermeegg}) with the state being AL (above this critical temperature $\ket{\Phi}$ experiences SD as well).

In considering a selection of familiar states for these two special cases of symmetric and anti-symmetric coupling, we find that the same sort of initial states can have very different entanglement dynamics when the spacing of the atoms is altered.  For the selection of initial states considered, the symmetric coupling, where $\Delta w = n \lambda/2$, is better for generating entanglement between the atoms, and for protecting entanglement from sudden death.  However, these remain only special cases that reveal some facets of the differences in entanglement dynamics arising from a change in distance.  It remains to be seen what happens for a more general selection of distances.

\section{General separation\label{sec:Gen}}

To extend the results found in the previous section to arbitrary atomic separations we use two different approaches to solve the problem --- the resolvent approach as described in \cite{ASH2006} and numerical diagonalization of the Hamiltonian. The first approach gives an analytic solution but only yields sufficiently simple expressions to be of use for initial states that have at most one excitation among the atoms and field modes, meaning superpositions and mixtures of states belonging to the set $\{\ket{\phi_i}\}=\{\ket{eg00},\ket{ge00},\ket{gg10},\ket{gg01}\}$. Using the resolvent expansion of the Hamiltonian
\begin{equation}
e^{-i\hat{H}t} = \int{\frac{dE e^{-iEt}}{E-\hat{H}+i\eta}}
\label{res}
\end{equation}
and expanding
\begin{multline}
	(E-\hat{H})^{-1} =(E-\hat{H}_0)^{-1}+(E-\hat{H}_0)^{-1}\hat{H}_I(E-\hat{H}_0)^{-1}\\
	+ (E-\hat{H}_0)^{-1}\hat{H}_I(E-\hat{H}_0)^{-1}\hat{H}_I(E-\hat{H}_0)^{-1}+... 
\label{pr}
\end{multline}
we can find an exact expression for the matrix elements $T_{ij}=\bra{\phi_i}(E-\hat{H})^{-1}\ket{\phi_j}$  by resumming the series expansion in (\ref{pr}). It can be seen that the action of the operator $\hat{A}=((E-\hat{H}_0)^{-1}\hat{H}_I)^2$ on the subsets $\{\ket{\alpha_i}\}=\{\ket{eg00},\ket{ge00}\}$ and $\{\ket{\beta_i}\}=\{\ket{gg10},\ket{gg01}\}$ gives a state in the same subspace. Say, $\op{A}(p^{(1)}_n\ket{\alpha_1}+q^{(1)}_n\ket{\alpha_2})=p^{(1)}_{n+1}\ket{\alpha_1}+q^{(1)}_{n+1}\ket{\alpha_2}$ and $\op{A}(p^{(2)}_n\ket{\beta_1}+q^{(2)}_n\ket{\beta_2})=p^{(2)}_{n+1}\ket{\beta_1}+q^{(2)}_{n+1}\ket{\beta_2}$, such that
\begin{align}
&\begin{pmatrix}
	p^{(1,2)}_{n+1}\\
	q^{(1,2)}_{n+1}
\end{pmatrix}=
A^{(1,2)}\begin{pmatrix}
	p^{(1,2)}_{n}\\
	q^{(1,2)}_{n}
\end{pmatrix} 
\end{align}
then,
\begin{equation}
A^{(1)}=\frac{2g^2}{(E-\omega_0)^{2}}\begin{pmatrix}
1 & e^{i\phi/2}\cos(\phi/2)\\
e^{-i\phi/2}\cos(\phi/2) & 1
\end{pmatrix}
\end{equation}
and $A^{(2)}={A^{(1)}}^\ast$.
Now consider the matrix element $T_{11}=\bra{eg00}(E-\hat{H})^{-1}\ket{eg00}$, by using the expansion in (\ref{pr}) this can be summed up to obtain
\begin{equation}
	\begin{split}
		T_{11} & =\frac{1}{4}\left(\frac{1}{E-\omega_0-2g\sin(\phi/4)}+\frac{1}{E-\omega_0+2g\sin(\phi/4)} \right. \\
 		& \left. +\frac{1}{E-\omega_0-2g\cos(\phi/4)}+\frac{1}{E-\omega_0+2g\cos(\phi/4)}\right) \\
		& =T_{22}=T_{33}=T_{44}
	\end{split}
\end{equation}
Similarly,
\begin{align}
	T_{12} &=\frac{e^{-i\phi/2}}{4}\left(\frac{1}{E-\omega_0-2g\cos(\phi/4)}+\frac{1}{E-\omega_0+2g\cos(\phi/4)} \right. \nonumber \\
	& \left. -\frac{1}{E-\omega_0-2g\sin(\phi/4)}-\frac{1}{E-\omega_0+2g\sin(\phi/4)}\right) \nonumber \\
	&=T_{21}^\ast=T_{34}^\ast=T_{43} \\
	T_{13} & = \frac{e^{-i\phi/4}}{4}\left(\frac{1}{E-\omega_0-2g\cos(\phi/4)}-\frac{1}{E-\omega_0+2g\cos(\phi/4)} \right. \nonumber \\
	& \left. +\frac{i}{E-\omega_0-2g\sin(\phi/4)}-\frac{i}{E-\omega_0+2g\sin(\phi/4)}\right) =T_{31}^\ast \nonumber \\
	&=T_{41}=T_{23}^\ast=T_{32}=T_{24}e^{-i\phi}=T_{42}^\ast e^{-i\phi}
\end{align}
From those matrix elements we obtain the matrix elements of the propagator as in (\ref{res}):
\begin{align}
	U_{11}&=\frac{e^{-i\omega_0t}}{2}\left(\cos{K_1t}+\cos{K_2t}\right) = U_{22}=U_{33}=U_{44}\\
	U_{12}&=\frac{e^{-i\omega_0t}e^{-i\phi/2}}{2}\left(\cos{K_1t}-\cos{K_2t}\right) = U_{43}\\
	U_{21}&=\frac{e^{-i\omega_0t}e^{i\phi/2}}{2}\left(\cos{K_1t}-\cos{K_2t}\right) = U_{34}\\
	U_{13}&=\frac{e^{-i\omega_0t}e^{-i\phi/4}}{2}\left(-i\sin{K_1t}+\sin{K_2t}\right) = U_{41} \nonumber \\
	 & =U_{32} = U_{24}e^{-i\phi}\\
	U_{31}&=\frac{e^{-i\omega_0t}e^{i\phi/4}}{2}\left(-i\sin{K_1t}-\sin{K_2t}\right) = U_{14} \nonumber \\
	& =U_{23}=U_{42}e^{i\phi}
\end{align}
where $K_1=2g\sin{\frac{\phi}{4}}$ and $K_2=2g\cos{\frac{\phi}{4}}$.

It can be seen as a general observation that for any initial state that lies in the single excitation subspace the time evolved atom-atom density matrix has the form
\begin{multline}
	\rho=\rho_{11}\ket{eg}\bra{eg}+\rho_{22}\ket{ge}\bra{ge}+\rho_{12}\ket{eg}\bra{ge} \\ 
	+\rho_{12}^\ast\ket{ge}\bra{eg}+\rho_{33}\ket{gg}\bra{gg}
\end{multline}
 \cite{JingLuFicek2009}.
For this general form of the density matrix the atom-atom concurrence simplifies to $C=2|\rho_{12}|$, meaning for a state in the single excitation subspace the entanglement of the two atoms has a dynamics that is either AL or DI (since $\rho_{12}$ is analytic and can never vanish over a finite non-zero length of time).

Using the propagator for the one-excitation subspace, we can examine the analytic expressions for entanglement dynamics of a few initial states.  We begin with the state $\ket{eg00}$.  The time evolved reduced density matrix of the two atoms after tracing out the field modes is obtained as
\begin{multline}
	\rho_{12}=\left|U_{11}\right|^2\ket{eg}\bra{eg}+\left|U_{21}\right|^2\ket{ge}\bra{ge}+U_{11}U_{21}^\ast\ket{eg}\bra{ge}\\
	+U_{11}^\ast U_{21}\ket{ge}\bra{eg}+\left(\left|U_{31}\right|^2+\left|U_{41}\right|^2\right)\ket{gg}\bra{gg}
\end{multline}
This gives a concurrence for the two qubits as
\begin{align}
C(\phi,t) = \frac{1}{4}\left|\cos(2K_1t)-\cos(2K_2t)\right|
\label{ceg00}
\end{align}
It can be seen from the above expression that, unless $\phi=\pi$, the atoms are entangled except at a discrete, periodic set of instants given by $t_d=m\pi/(K_1-K_2)$, $m\epsilon I$ where the state is separable (i.e., ``dead'').  As the separation parameter $\phi$ approaches $\pi$, $K_1$ approaches $K_2$ and the death time approaches infinity. Thus for antisymmetric coupling, there is no entanglement generated at all, as the situation corresponds to having the two atoms interacting with two separate transformed modes.  Thus, this special case is uncharacteristic of the qualitative features of the entanglement dynamics for general separations, and for any other separation the atoms will become entangled, if perhaps on a long timescale.
\begin{figure*}[ht]
\subfloat[]{\label{eg001}\includegraphics[width=0.45 \textwidth]{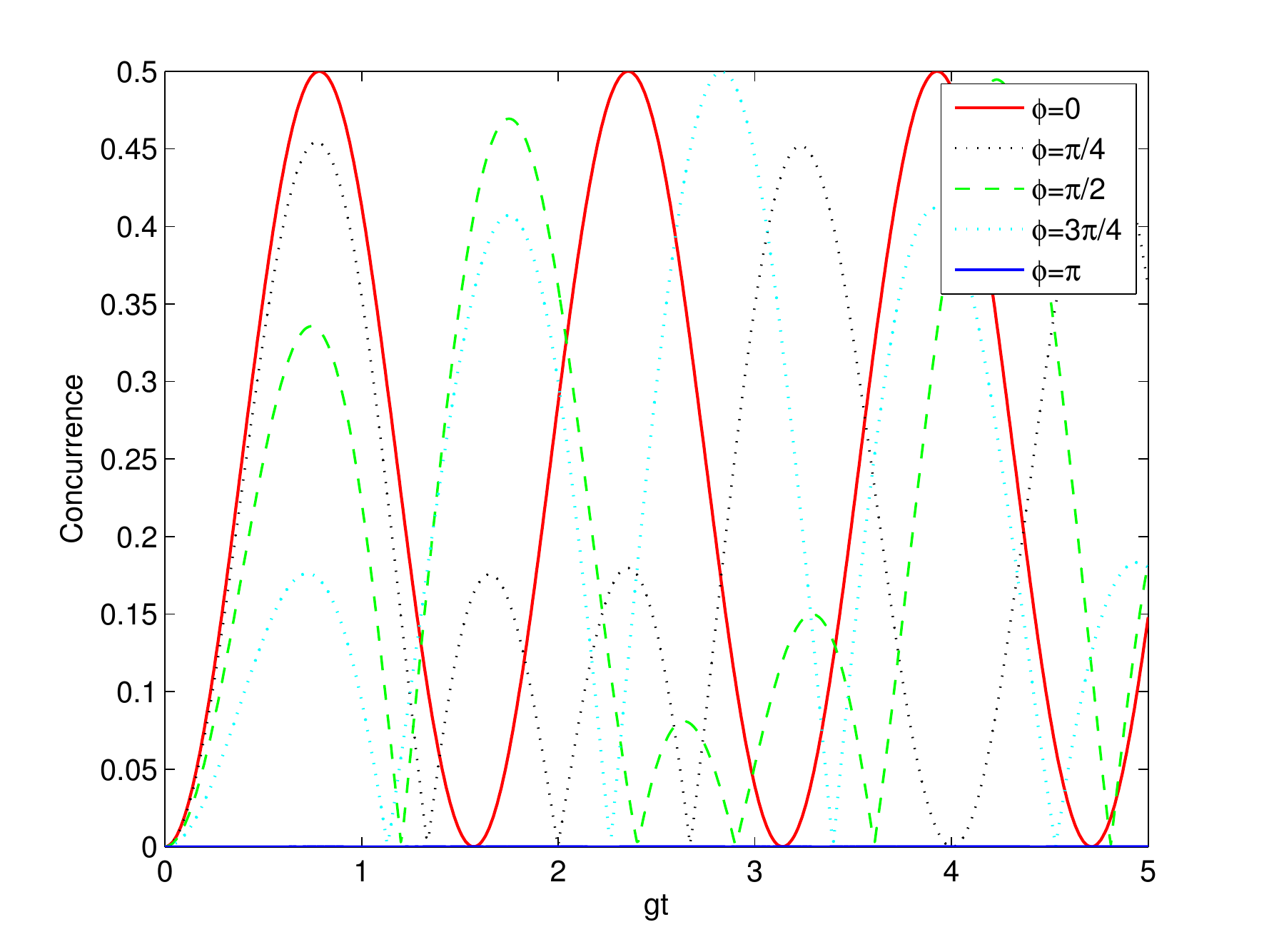}}
\subfloat[]{\label{eg002}\includegraphics[width=0.45 \textwidth]{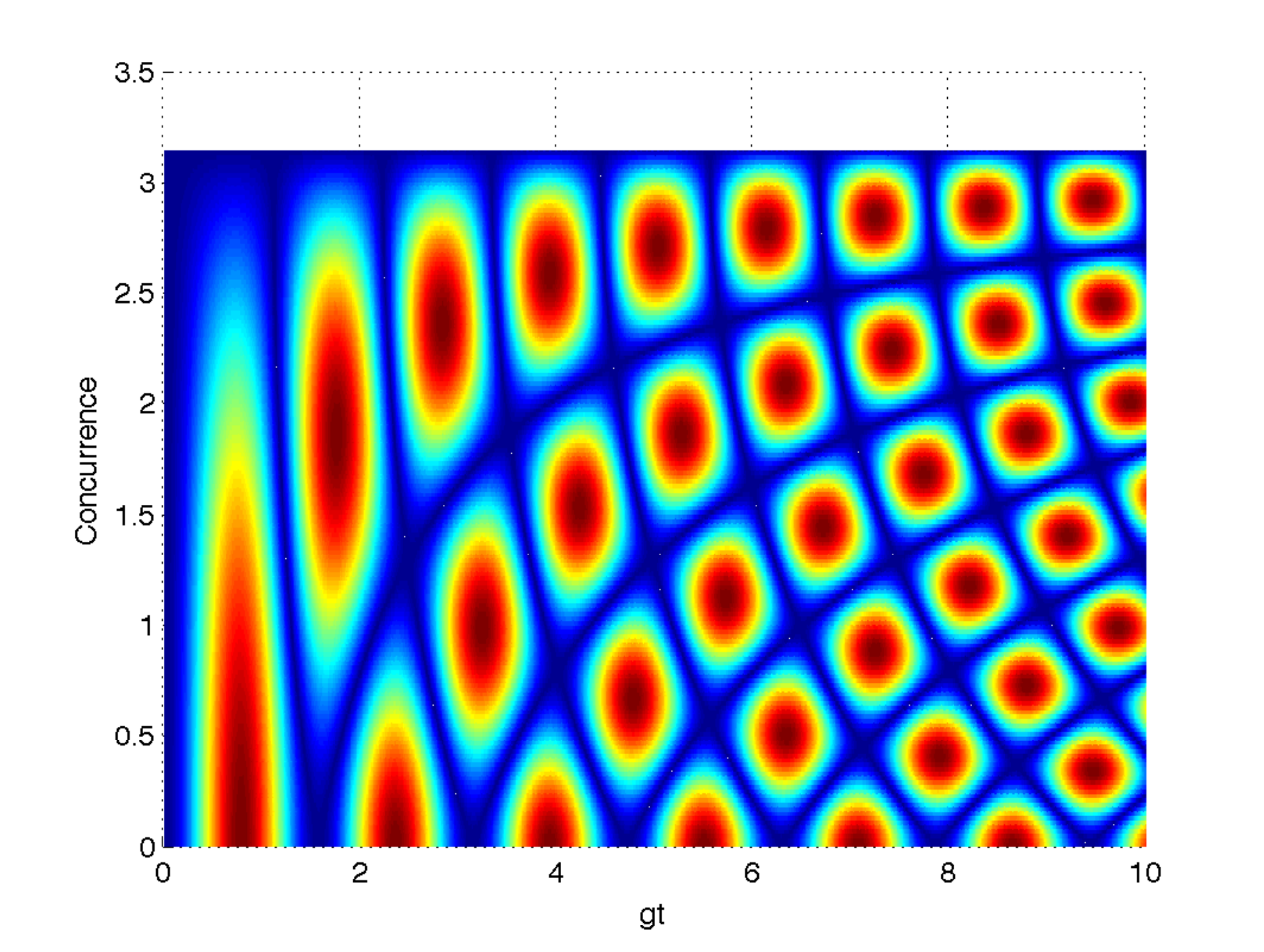}}\\
\caption{Entanglement dynamics for $\ket{eg00}$, DI behavior for all values of $\phi$}
\label{eg00}
\end{figure*}

For the initial state $\ket{gg10}$ the atom-atom entanglement is 
\begin{align}
C(\phi,t) = \frac{1}{2}\left(|\sin^2(K_1t)|+|\sin^2(K_2t)|\right).
\end{align}
In this situation it can be seen that the entanglement remains AL except for separations where $\tan(\phi/4)$ is rational, where the entanglement dynamics is DI.  This latter case includes symmetric and antisymmeric couplings, where the concurrence oscillates as $C(0,t)=1/2|\sin^2(2gt)|$ and $C(\pi,t)=|\sin^2(\sqrt{2}gt)|$ respectively.
\begin{figure*}[ht]
\subfloat[]{\label{gg101}\includegraphics[width=0.45 \textwidth]{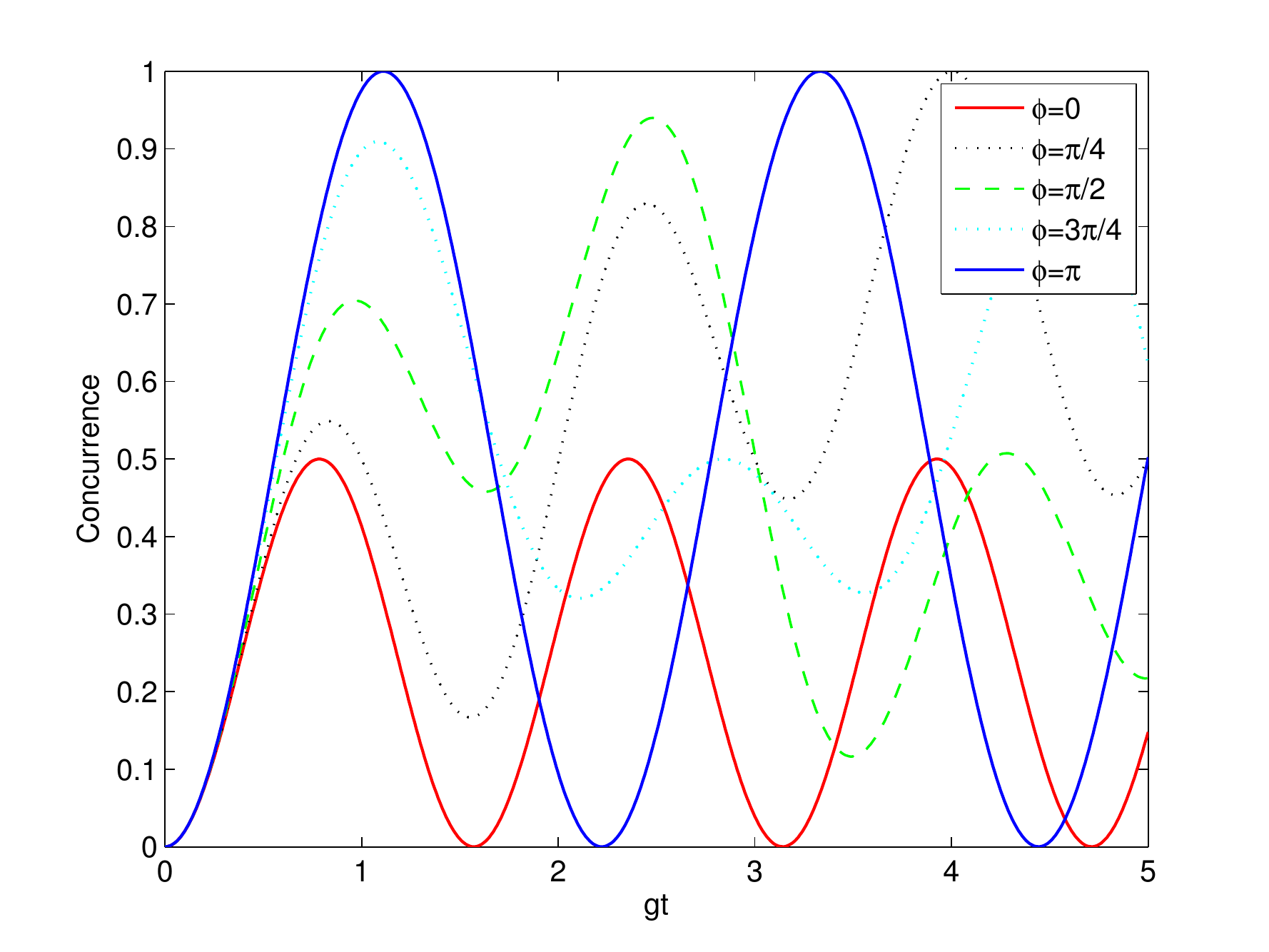}}
\subfloat[]{\label{gg102}\includegraphics[width=0.45 \textwidth]{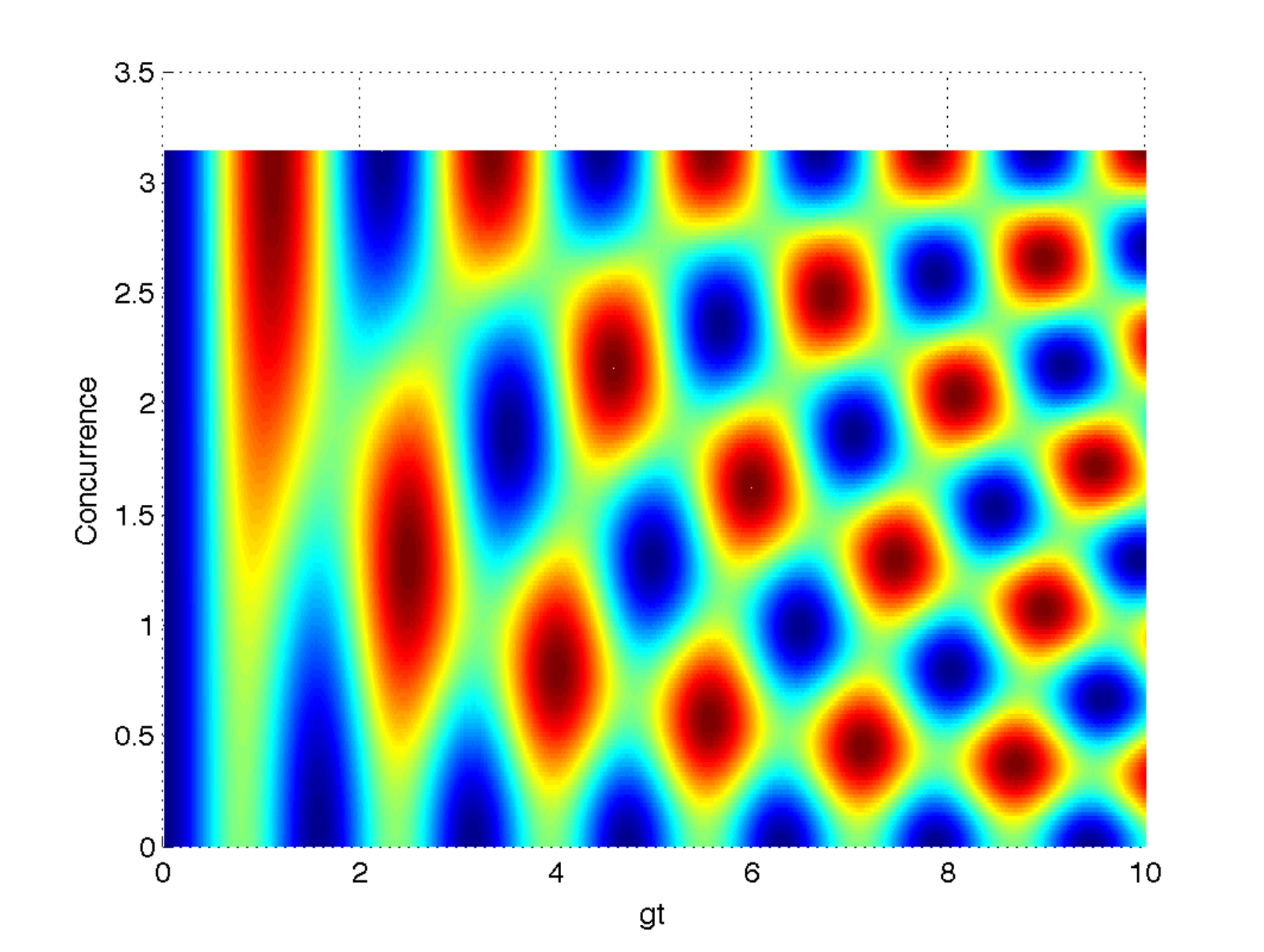}}\\
\caption{Entanglement dynamics for $\ket{gg10}$, Al/DI behavior for all separations}
\label{gg10}
\end{figure*}
Here we see that the special cases describing states with high symmetry are not at all representative of the generic qualitative features for arbitrary separation.
For the initially entangled atomic state $\frac{1}{\sqrt{2}}(\ket{eg}+\ket{ge})\ket{00}$, the concurrence is given as $C(\phi,t)=\cos^2(K_1t)\cos^2(\phi/4)+\cos^2(K_2t)\sin^2(\phi/4)$, again we can see that entanglement is AL 
unless $\tan(\phi/4)$ is a rational number.
\begin{figure*}[ht]
\subfloat[]{\label{egge001}\includegraphics[width=0.45 \textwidth]{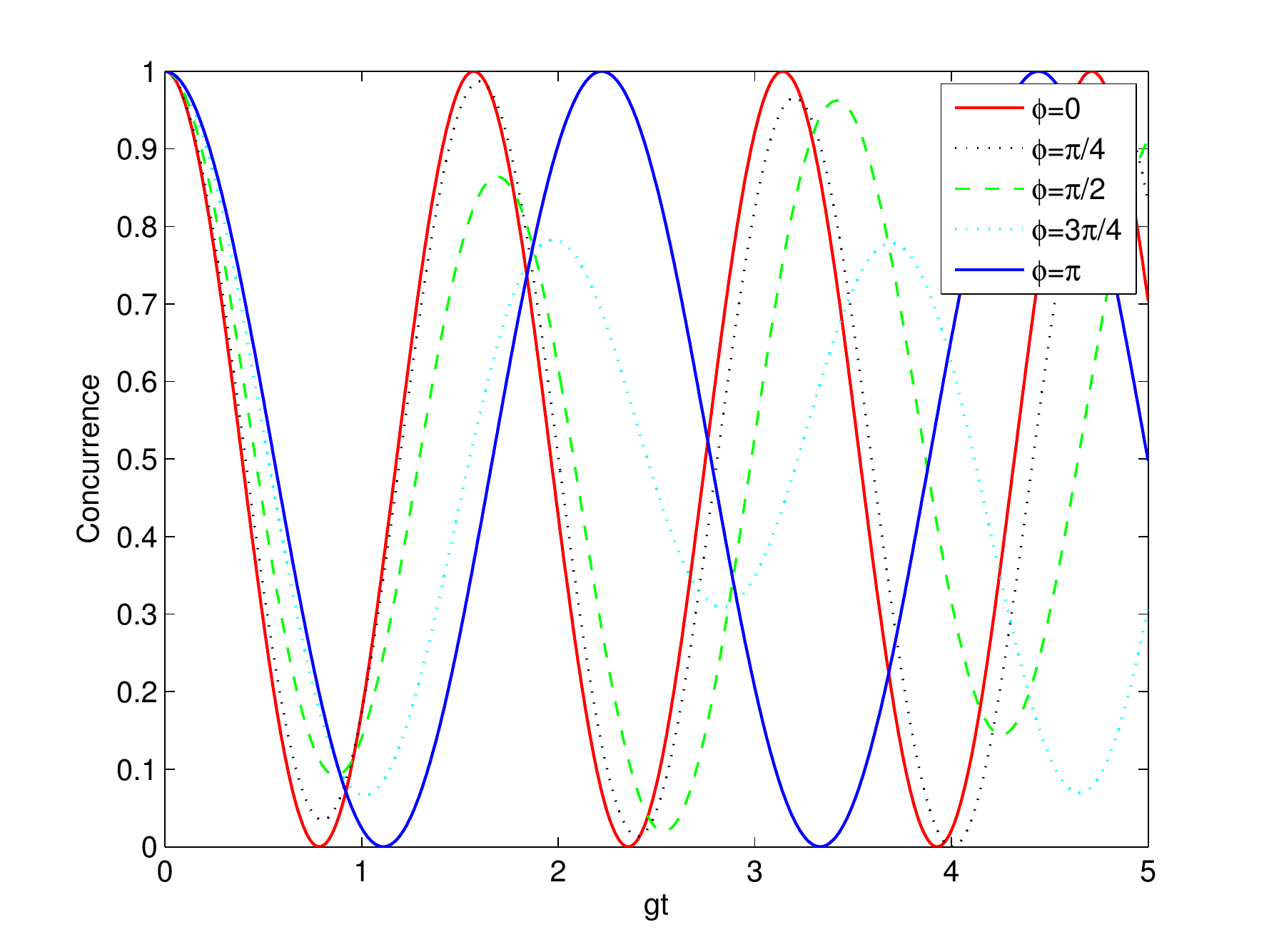}}
\subfloat[]{\label{egge002}\includegraphics[width=0.45 \textwidth]{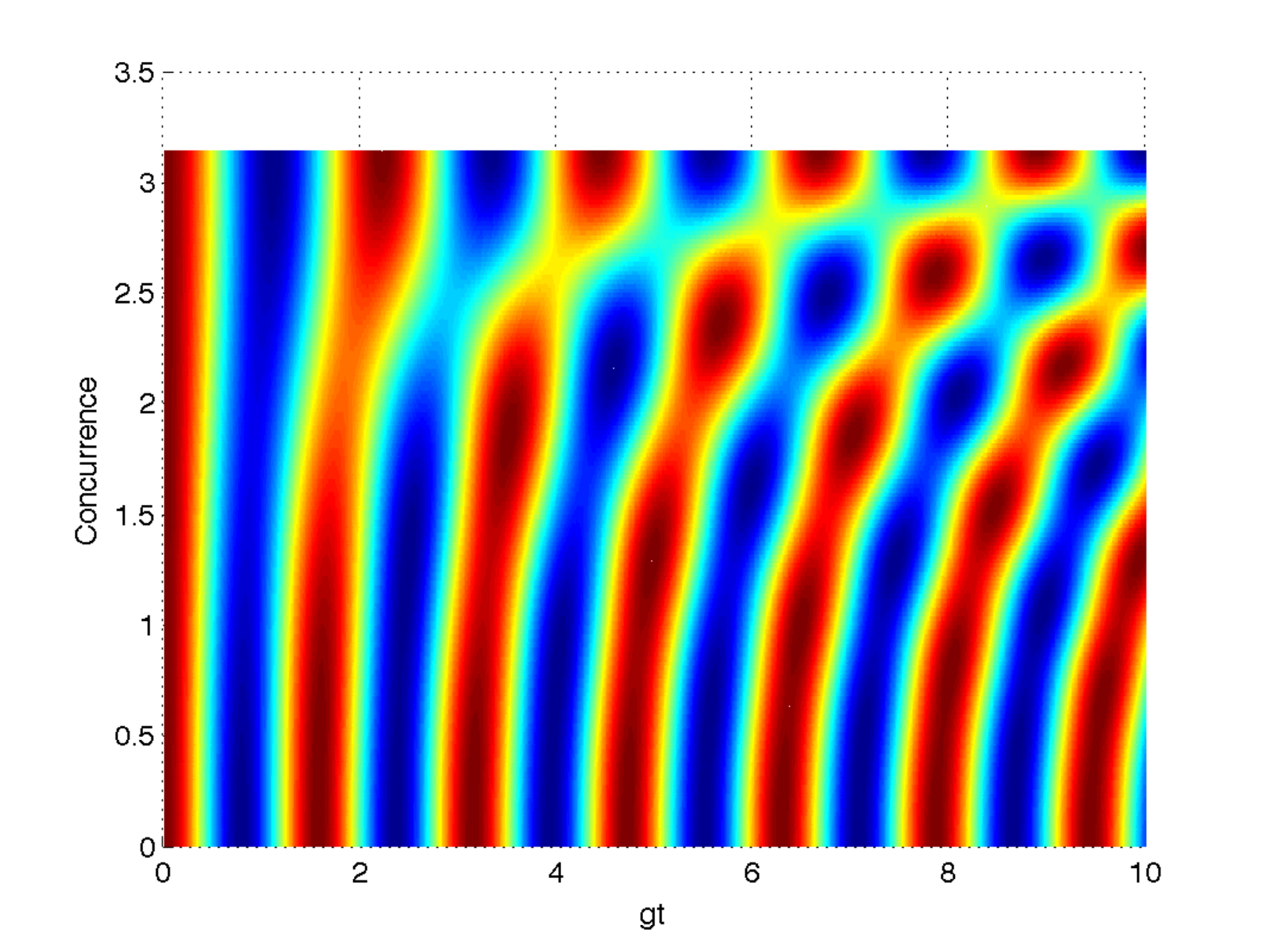}}\\
\caption{Entanglement dynamics for $\frac{1}{\sqrt{2}}(\ket{eg}+\ket{ge})\ket{00}$, AL/DI behavior for all values of separations}
\label{egge00}
\end{figure*}

Restricting to the single excitation subspace, one might also seek to find the approximate concurrence for thermal and coherent states with low average photon numbers (keeping only one excitation in the field modes) interacting with $\ket{gg}$; however, one can find that the resulting values are of the same order as two-excitation terms that have been neglected, so this approach is not useful in that case. 

\begin{figure*}[ht]
\subfloat[Initial state $\ket{\Phi}$ interacting with a thermal state with average number of photons $n_{th}=0.2$]{\label{sdth}\includegraphics[width=0.45 \textwidth]{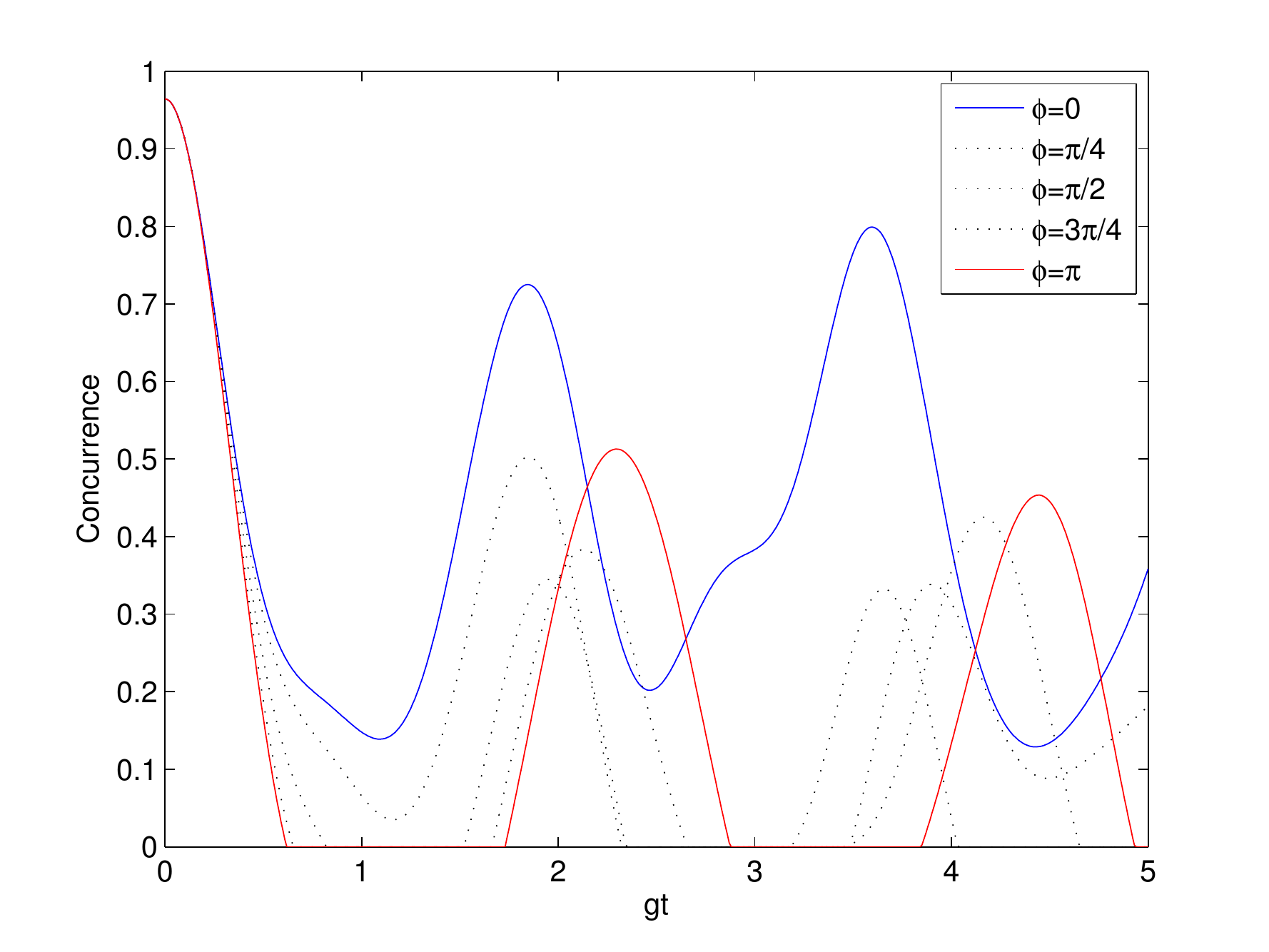}}
\subfloat[$\ket{\Phi}$ interacting with a two-mode squeezed state with $\xi_{sq}=0.1$]{\label{sdtmss}\includegraphics[width=0.45 \textwidth]{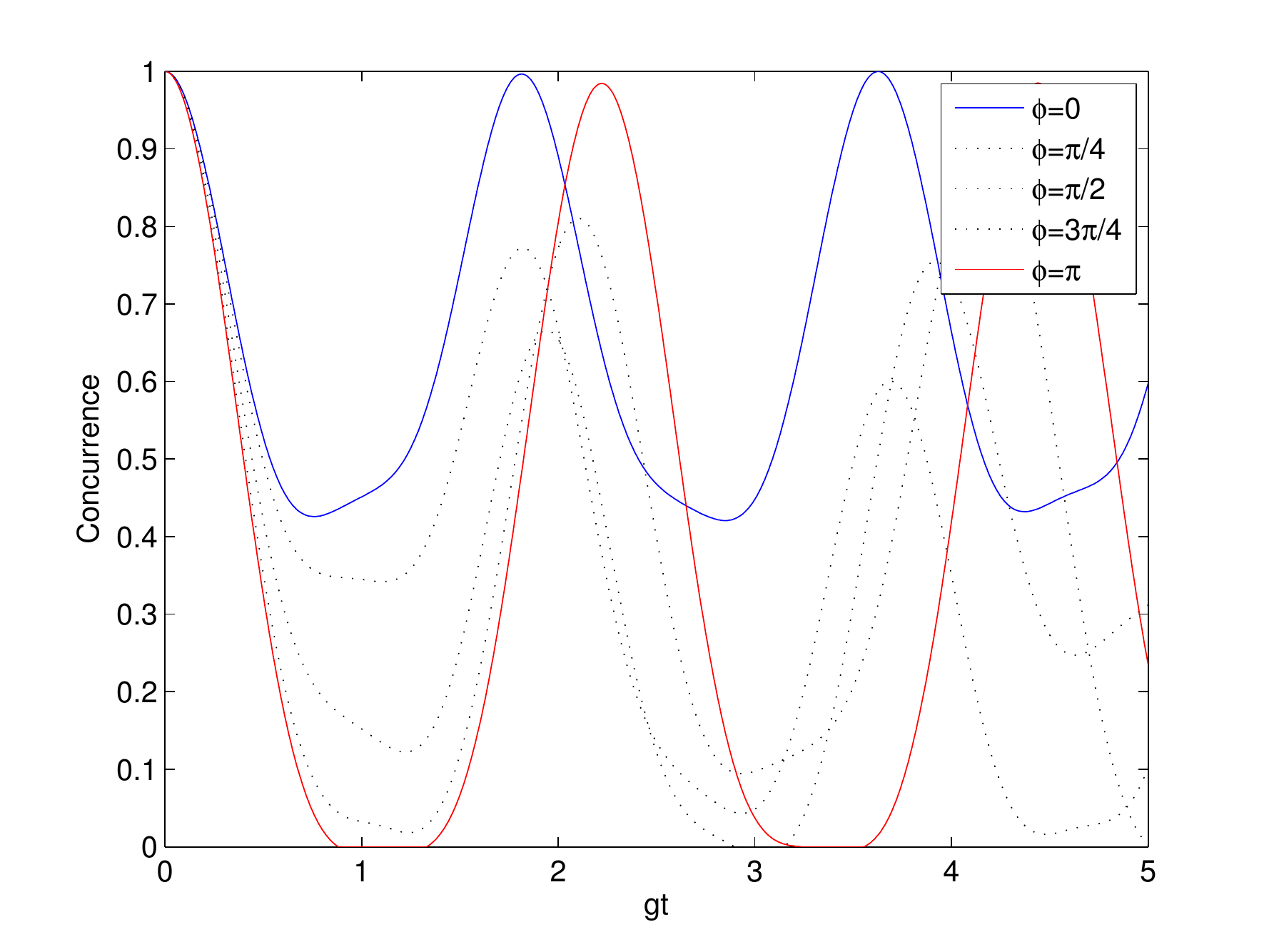}}\\
\caption{Entanglement dynamics for $\ket{\Phi}$, AL behavior for symmetric coupling, SD for all the other separations}
\label{SDgen}
\end{figure*}

\begin{figure*}[ht]
\subfloat[$\ket{eg22}$ has DI for symmetric coupling ($\phi=0$) and no entanglement for antisymmetric coupling ($\phi=\pi$), SD for all the intermediate separations]{\label{eg22}\includegraphics[width=0.45 \textwidth]{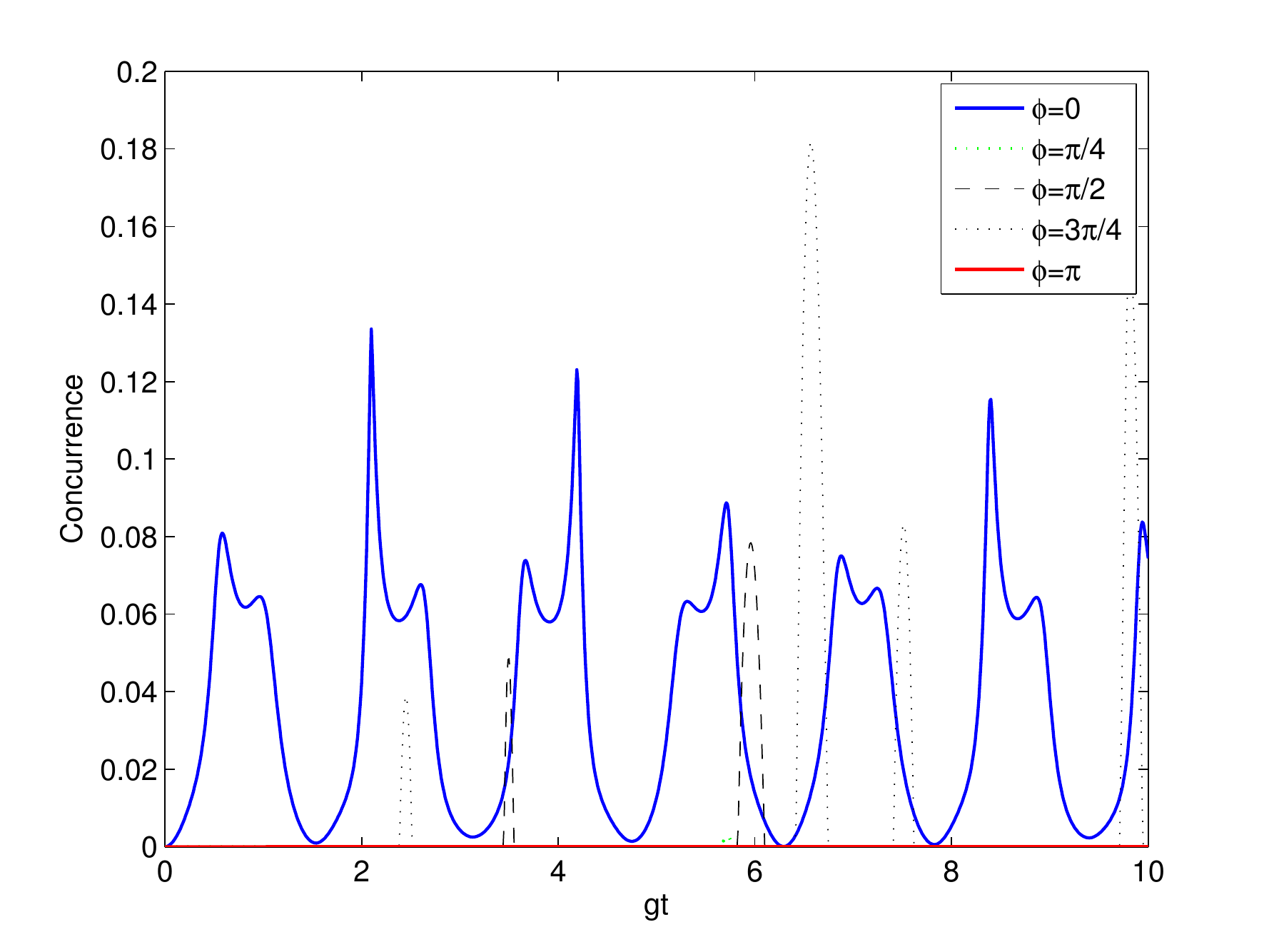}}
\subfloat[$\ket{ee11}$ shows no entanglement generation for the symmetric and anti-symmetric coupling, SD otherwise]{\label{ee11}\includegraphics[width=0.45 \textwidth]{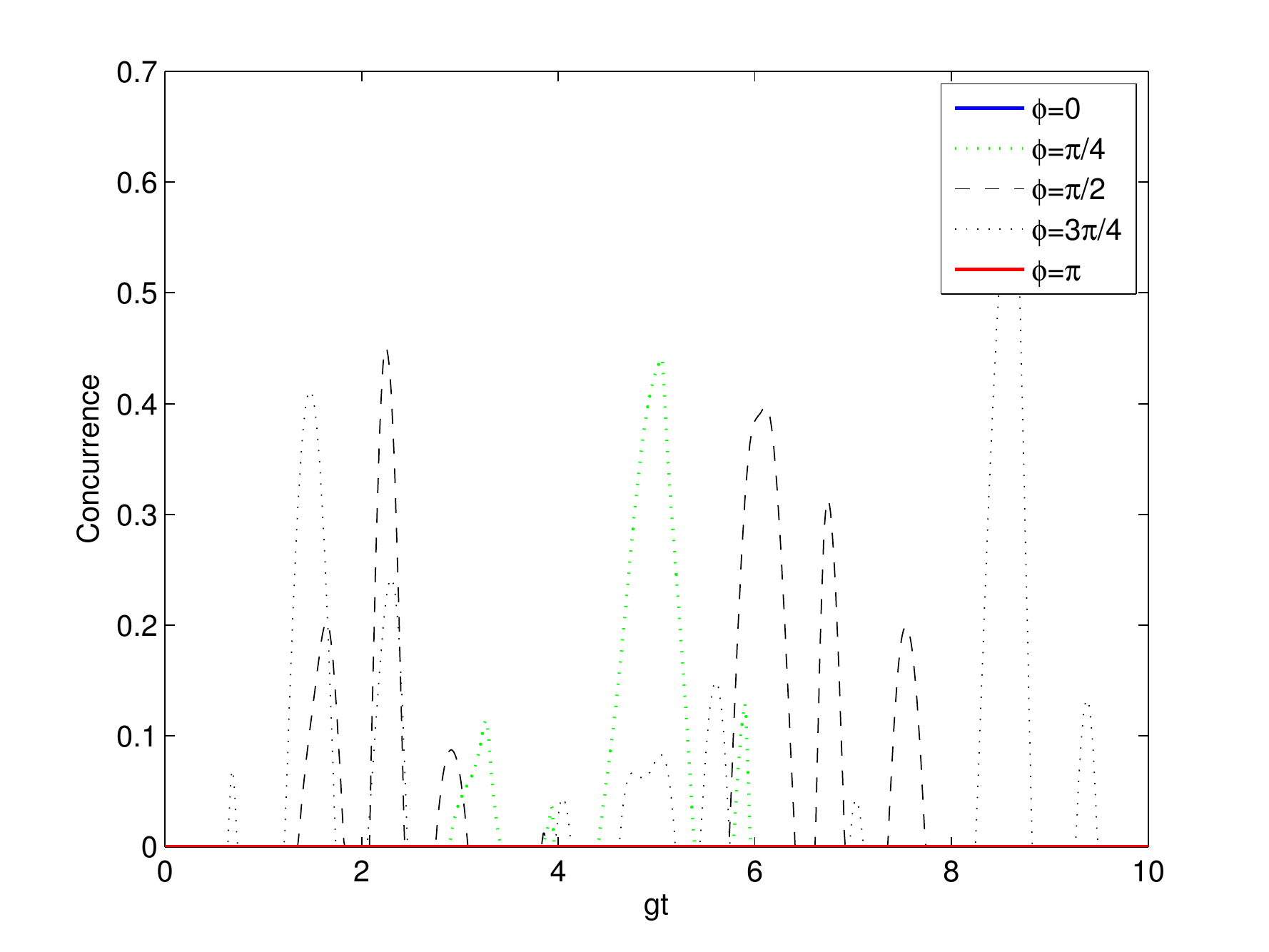}}\\
\caption{Entanglement dynamics for Fock states with general separations}
\label{fockgen}
\end{figure*}

To look at the entanglement dynamics for other initial states we numerically diagonalize the Hamiltonian, restricting to a small photon number approximation. We find that SD is generally observed for most distances with the initial states we have discussed in the preceding section. In many cases, however, the special cases of symmetric and antisymmetric coupling show qualitatively different dynamical behavior. For example, if we look at the entanglement dynamics for the maximally entangled state $\ket{\Phi}$ interacting with a thermal state in the SC model, we had seen that the entanglement was protected upto some critical temperature, while there was SD for the AC model. Upon extending this to arbitrary separation values we find that for all separations where $\phi \ne 0$ the entanglement always shows SD for all temperatures as shown in Fig. \ref{SDgen}. Similarly, for $\ket{\Phi}$ interacting with a two-mode squeezed state one has SD except for symmetric coupling or $\phi=0$.
As we had discussed in the previous section for the case of an initial EM field state diagonal in the Fock basis in the TC model, if both the atoms are excited there is no entanglement generation and for the atoms in $\ket{eg}$ the entanglement dies only for an instant. Considering the Fock state $\ket{n_N,n_N}$ in the SC model translates to the state $\op{\rho}_{nn}$ in the TC model which is diagonal in the Fock basis, hence we see no entanglement generation for $\ket{ee}$ and DI for $\ket{eg}$. On the other hand, we had also shown that for such a state there is no entanglement generation in the AC situation. We use numerical approach to consider the entanglement behavior for intermediate separations. It can be seen from Fig.\ref{fockgen} that the intermediate distances have sudden death for the case of $\ket{egnn}$, even if one deviates slightly from the symmetric coupling $\phi=0$. The state $\ket{eenn}$ shows entanglement generation for asymmetric couplings.


While examining the special cases with high symmetry in the setup it is useful to show the over-riding effect of symmetry on the entanglement dynamics of this system, computing the results for arbitrary distances suggests that these special cases are not representative in terms of the general qualitative behavior of entanglement.  
As we have discussed above, in many cases the dynamics falls in one qualitative class almost everywhere on the interval 
of asymmetrically coupled cases ($\phi\ne0$ or $\pi$) but in a different class on one or both boundaries.  Moreover, of the three qualitative 
classes we have discussed here, all are possible at every distance, given the correct initial state.  An analysis of linear oscillators coupled to a common vacuum field (consisting of a continuum of modes) by Lin and Hu found that there exists an entanglement distance, so that oscillators with a separation less than that distance will be asymptotically entangled, while initially entangled oscillators spaced further will undergo SD \cite{Lin09}.  In contrast, an analysis of two two-level atoms interacting with a common EM thermal field (with a continuum of modes) found \cite{Ndipole} that all initial states will eventually undergo SD for any non-zero separation.  In this two-mode model we find no evidence of an ``entanglement distance'' which determines a change in the qualitative features of entanglement dynamics.  For each of initial states considered, the qualitative class into which the entanglement dynamics falls is the same for almost all distances, with departures only on a set of measure zero.  For a thermal field, which is the environment considered in \cite{Lin09,Ndipole} (the former considers specifically the vacuum), SD occurs at almost all distances.  This bears some resemblance to the result of \cite{Ndipole}, though the non-dissipative nature of this model ensures revival and, in this regard, is qualitatively quite different.

\section{Summary and Discussion\label{sec:conclusion}}

We have analyzed the entanglement dynamics in a model consisting of
two two-level atoms and two electromagnetic field modes for a
variety of familiar field states and classified the various cases in
terms of different dynamical behavior. 
One aim of this analysis is to get a sense of the variety of entanglement dynamics that can arise from different atomic separations in a case where two atoms interact with a shared EM field.  It is useful to examine this question in a very simple model that can be solved with a minimum of approximations (using which can
lead to unphysical effects) and understood in detail. We have argued
that there is no non-trivial distance dependence in a single mode
model, and, therefore, a two-mode model represents the simplest case
for the study of distance dependence in the entanglement dynamics.
On the other hand, including a continuum of modes leads to dissipation and misses the variety of dynamical behavior that can be seen with a few number of modes. In particular, the effect of initial state is largely washed out.
We have first studied two arguably extreme cases out of the class of
two-mode Hamiltonians that can arise, one in which the two modes are
symmetrically coupled (SC) to the second atom, and one where the
two modes have asymmetric coupling (AC).  A useful insight in
understanding these models is that the atomic dynamics in the SC
model correspond exactly to the dynamics for a Tavis-Cummings model with a single
field mode symmetrically coupled (TC) to both atoms with a suitable
mapping of the field state, while the atomic dynamics for the AC
model correspond exactly to the dynamics for a double Jaynes-Cummings
(DJC) model, made up of two isolated subsystems each with one atom
coupled to one mode, under a proper transformation of the field
state. These mappings help one understand the significant differences
in behavior in the two seemingly similar models, giving a window into
how symmetry introduced by specific atomic separation can affect entanglement dynamics.
Another significant implication of the mapping between the SC and
TC models comes from the fact that the mapping of initial field
states between those models is many-to-one, this shows us entire classes of field states for the
SC model that will give exactly the same atomic dynamics.

In examining the dynamical generation of entanglement  from an
initially separable atomic state, we find quite
a marked contrast between the SC and AC models.  While
entanglement generation is a relatively common feature, present for a
variety of field states, in the SC model, it is comparatively much
more rare in the AC model. This difference is not so surprising
however, if one views it in terms of the mappings we have introduced
to the other models.  When considering that the AC maps to the DJC
model, where the two subsystems are isolated, one would expect
entanglement generation to be relatively rare; it can only exist in
cases where the field state is mapped to an entangled state whose
entanglement can then be transferred to the atoms.  With the SC, by
contrast, we have a mapping to the TC model where a single shared
field mode can readily introduce entanglement between the two atoms.

These models (TC and DJC) have previously been studied for some of the cases that we have looked at as well. For example, Tessier et al in \cite{TessierDeutsch2003} had also analyzed the the initial state $\ket{een}$ and coherent state in the TC model. Also, interaction of atomic states $\ket{ee}$, $\ket{eg}$ and $\ket{gg}$ with a thermal field was studied in the context of thermal generation of entanglement in \cite{KimKnight2003}. We extend the conclusion of these results more generally to the set of initial field states that are diagonal in the Fock basis. We find that for a field state diagonal in the Fock basis there is no entanglement generation in $\ket{ee}$, DI for $\ket{eg}$ and for $\ket{gg}$ an initial field state with a high component of $\ket 1\bra 1$ would generate more entanglement in general. Other than that we study the case of an initially entangled atomic state interacting with a thermal field and find that the entanglement is preserved for low temperatures and the incidence of sudden death occurs only after a certain critical temperature.
The DJC model was previously studied from the viewpoint of transfer of entanglement phenomenon in \cite{YonacYuEberly2006, YonacYuEberly2007} and more generally by \cite{Chan2009} for the case of an entangled atomic state interacting with a vacuum field and transferring correlations via the atom-field entanglement.
We look at the phenomenon for the reversed situation, where one starts with a maximally correlated field state (TMSS) and looks at the entanglement generated in the atomic subsystem. Since the field state does not remain gaussian as the system evolves, we resort to a low temperature approximation in order to quantify the entanglement. We also see that the transfer is maximum for an optimal squeezing parameter value such that the field state approximates a maximally entangled qubit state.
As a potentially useful phenomenon we observe cases where we find that the entanglement always stays alive once generated or stays protected if already present. In particular, for the case of a squeezed state interacting with an initially entangled atomic state $\ket{\Phi}$ we see that the qubits end up in an almost dark state and stay almost maximally entangled for all times.
As a natural question one may ask what would be the extension of these two extreme cases to intermediate separations and how does the dynamics on one extreme change into a different behavior on the other. We look at the general separations using the resolvent method for states in the single excitation subspace to find the concurrence as a function of separation and time. We see that these states can only exhibit AL/DI behavior based on the general form of the atom-atom density matrix as noted in \cite{JingLuFicek2009}, in particular it can be seen explicitly how the behavior changes for the symmetric and anti-symmetric couplings. We can look at all the other cases using numerical analysis and we find that for all the cases considered there is no critical separation behavior as seen in the continuum cases \cite{Lin09,Ndipole}. For example, situations where entanglement stays protected in the symmetric coupling case (thermal state and squeezed state) shows SD as soon as there is any asymmetry in the coupling.

We have chosen in this analysis to try to isolate the effect of atomic separation on entanglement dynamics from the position-dependent effects arising from boundary conditions.  For this purpose, we have assumed in Sec.~\ref{sec:sysham} an atom-field coupling that depends on the coordinate separating the atoms only by a phase factor.  For clarity we have supposed our two field modes are traveling-wave modes in free space, however a more experimentally relevant situation would be a toroidal resonator, where there is rotational invariance in the azimuthal direction that satisfies our requirements.  In this case the two modes of interest would be two resonant, counter-propagating whispering-gallery modes.  Because strong coupling between an atom and a whispering-gallery modes of a microtoroidal resonator has observed experimentally \cite{KimbleTorus,DanAlton}, there is the possibility of experimentally probing quite directly the model we have considered.  However, in the experimental system there will be dissipative dynamics arising from emission into other modes outside the resonator, so a detailed comparison would require either including the dissipative effects in the theoretical model or a restriction to the case of sufficiently strong coupling to the resonator modes and early times that the dissipation could be neglected.

Our analysis of special cases of the entanglement dynamics arising
in two atoms interacting with two modes suggests a wide variety of
different behaviors can arise, with qualitative features of the
dynamics changing entirely between the two special cases considered, even with the same
initial field state.  This suggests that understanding the distance
dependence of entanglement dynamics for multiple atoms interacting
with a common field will be quite important for predicting even the
qualitative features that may arise. Furthermore, if one has the
practical goal of dynamically generating entanglement or protecting
entanglement once generated, the special cases we have considered
suggest that the ability to achieve these goals will be greatly
impacted by the separation of the atoms.  Having gotten some sense of
the variation in behavior that can arise, we quantify the entanglement dynamics over a range of
atomic separations rather than just special or extreme cases that highlight the importance of symmetry factors in entanglement dynamics.

On this point it would be interesting to compare the exact results of the present model with the general statements on classifying the entanglement behavior \cite{Joynt2011} based on geometrical and topological properties of quantum state spaces \cite{GeometryQS}.

\acknowledgments
This research is supported in part by grants from the NSA-Laboratory for Physical Sciences, the DARPA-QuEST program (DARPAHR0011-09-1-0008) and  the NSF-ITR program (PHY-0426696). Results here were reported by BLH at the program on Quantum Open Systems at KITP-China (Beijing) supported by the Project of Knowledge Innovation Program (PKIP) of the Chinese Academy of Sciences, Grant No. KJCX2.YW.W10.

\end{document}